\documentstyle[11pt,aaspp4]{article} 

\begin{document}

\title{THE VIRGO CLUSTER DISTANCE FROM 21 CM-LINE WIDTHS}

\author{Martin Federspiel, G.A.~Tammann}
\affil{Astronomisches Institut der Universit\"at Basel,
            Venusstr.~7, CH-4102 Binningen, Switzerland}
\authoremail{martin@sirius.astro.unibas.ch}

\author{Allan Sandage}
\affil{The Observatories of the Carnegie Institution of Washingtion,
813 Santa Barbara Street, Pasadena, CA 91101}

\begin{abstract}
The distance of the Virgo cluster is derived in the $B$ band from the
21$\,$cm-line width-absolute magnitude relation. The latter is
calibrated using 18 spirals with Cepheid distances mainly from HST.
The calibration is applied to a {\sl complete} sample of non-peculiar
spirals with $i>45^{\circ}$ and lying within the optical (n=49) or
X-ray (n=35) contour of the cluster, resulting in a mean cluster
distance of $(m-M)^0=31\fm 58\pm 0\fm 24$ (external error) or
$20.7\pm 2.4$ Mpc. The mean distance of subcluster A is $0\fm
46\pm 0\fm 18$ smaller than that of subcluster B, but the individual
distances of the members of the two substructures show considerable
overlap. Cluster spirals with $30^{\circ}<i<45^{\circ}$ yield almost
as good distances as more inclined galaxies. H$\,$I-truncated galaxies
are overluminous by 0\fm 8 at a given line width. The distance modulus
is corrected by $-0\fm 07$ for the fact that cluster members have
lower H$\,$I-surface fluxes and are redder in $(B-I)$ at a given line
width than the (field) calibrators. Different sources of the $B$
magnitudes and line widths have little effect on the resulting
distance. Different precepts for the internal-absorption correction
change the result by no more than $\pm 0\fm 17$. The individual
distances of the cluster members do not show any dependence on
recession velocity, inclination, Hubble type or line width. The
dependence on apparent magnitude reflects the considerable depth
effect of the cluster. The adopted distance is in good agreement with
independent distance determinations of the cluster.

Combining the cluster distance with the corrected cluster velocity of
$1142\pm 61$ km s$^{-1}$ gives $H_0=55\pm 7$ km s$^{-1}$
Mpc$^{-1}$ (external error). If the Virgo cluster distance is inserted
into the tight Hubble diagram of clusters out to 11$\,$000 km s$^{-1}$
using {\sl relative} distances to the Virgo cluster one obtains a
global value of $H_0=57\pm 7$ km s$^{-1}$ Mpc$^{-1}$.
\end{abstract}


\section{Introduction}

The correlation of the rotational velocity and the luminosity of a
galaxy was first exploited by \"Opik (1921, 1922) in papers well ahead of
their time. Subsequently Gouguenheim (1969) suggested 21$\,$cm line
widths to be used as a measure of the rotational velocity and hence of
the galaxy luminosity. The ensuing 21$\,$cm-line width-absolute
magnitude relation of the form
\begin{equation} 
M= a \log w +b 
\end{equation}
is now known as \lq\lq Tully-Fisher\rq\rq (TF) relation (Tully \& Fisher 1977).

The first application of the TF method to the Virgo cluster gave a
distance modulus of $(m-M)=31.70\pm 0.08$ (Sandage \& Tammann 1976).
This relatively large value was confirmed by Kraan-Korteweg, Cameron
\& Tammann (1988) and, after correction of the distances of the local
calibrators, by Fouqu\'e et al.~(1990) using a {\sl complete} cluster
sample.  Incomplete cluster samples, however, have consistently given
significantly smaller TF distances (Aaronson, Huchra, \& Mould 1979,
Pierce \& Tully 1988).  This is now understood as the result of the
Teerikorpi cluster effect (Teerikorpi 1984, 1987, 1990, 1993; Sandage,
Tammann \& Federspiel 1995), which always leads to too small
distances if an incomplete set of cluster galaxies is used to define
the slope of the TF relation (cf.~Fig.~13 below) or of any other
relation between an observational parameter and the absolute magnitude
in the presence of intrinsic scatter.

Justification for a re-examination of the TF distance comes from
several facts. (1) The calibration of the TF relation can now be
based on an unprecedended sample of spiral galaxies with Cepheid
distances mainly from HST, and a complete sample of cluster spirals
can now objectively be selected from optical as well as X-ray surveys
(Sects.~2 and 5.3). (2) The resulting zero-point of the TF relation
is firm (Sect.~2.1) and the mean distance of a complete sample of
49 inclined Virgo spirals carries a correspondingly small systematic
and statistical error (Sect.~5). Largely independent data sets and
different correction procedures allow an assessment of the influence of
varying the input parameters (Sect.~6).
Several tests (Sect.~7) are used to demonstrate the internal
consistency of the individual galaxy distances. In Sect.~8 we investigate
the difference between the calibrators and the Virgo spirals in
H$\,$I surface brightness and $(B-I)$ color and its effect on the
derived distances. The spatial structure of the Virgo cluster
as outlined by the TF distances is illustrated in Sec.~9. The discussion 
in Sect.~10 compares the results with external data. The conclusion 
with respect to $H_0$ are presented in Sect.~11. 

\section{Sample Selection}

\subsection{Calibrators}

During the last few years the number of galaxies with Cepheid
distances has been growing significantly due to the capabilities of
the Hubble Space Telescope. For the first time there are enough
galaxies with Cepheid distances for a reliable calibration of the TF
relation. We use 18 spirals of Hubble types $T=2$ to 5 of which 12
have Cepheid distances determined with HST, and 4 have ground-based
Cepheid distances set out later in Table 2. Two late-type members
($T=7$) of the tight M$\,$101 group (NGC$\,$5204 and NGC$\,$5585)
without individual Cepheid distance were added to extend the range of
Hubble types of the sample. These two galaxies were assigned the
HST-Cepheid distance of M$\,$101. We consider the M$\,$81 and the
Sculptor group to be too spread in distance and do not include members
of these groups without individual Cepheid distances. NGC$\,$4496A,
which has a Cepheid distance (Saha et al.~1996a), is not used because
its total magnitude is disturbed by NGC$\,$4496B.

The zeropoint of the P-L relation of Madore \& Freedman (1991) which
we have used rests on the assumption that the true distance modulus of
LMC is $(m-M)^0=18\fm 50$. Eleven distance determinations based on
geometrical, physical and astronomical zeropoints give consistently
$(m-M)^0=18\fm 56\pm 0\fm 03$ shown in Table 1. If this value is taken
at face value, the derived Virgo cluster modulus would increase by
$0\fm 06\pm 0\fm 03$, and $H_0$ would be decreased by $\sim 3$\% from
the value we derive in Sect.~11. In addition, the majority of the Cepheid
distances of the calibrators in Table 2 are taken from the original
literature, i.$\,$e.~they are not corrected for the long-exposure
effect of the WFPC2 (Saha et al.~1996b; Hill et al.~1997). This effect
would increase the Cepheid distances of several galaxies by another
0\fm 05 and decrease $H_0$ by another $\sim 2$\%. On the other hand
Tanvir (1997) has pointed out a slight inconsistency of the $I$
magnitudes used by Madore \& Freedman (1991) affecting the Cepheid
distances determined from $V$ and $I$ magnitudes with HST. He proposes
$H_0$ to be larger by 5\% for this effect; since 13 out of 18
calibrators in Table 2 depend on $V$ and $I$ data from HST this
causes a 3-4\% increase of the adopted value of $H_0$ in Sect.~11.
This should approximately be balanced by a somewhat higher LMC
modulus in Table 1 and the long-exposure effect. In addition there 
is at least the suspicion that HST photometry in crowded fields
gives too {\sl bright} magnitudes (Sodemann \& Thomson 1997); for
the relevant Cepheids the effect is estimated to be $\sim 0\fm 05$ 
which leads to a corresponding underestimate of the distances (Saha
\& Labhardt 1997).

In a preprint C.S.~Kochanek (1997) has calculated Cepheid distances of
15 galaxies relative to LMC [again assumed to be at $(m-M)^0=18\fm
50$] by combining all available multi-color Cepheid data and assuming
that they are strictly on a homogeneous photometric system and have
equal weight. Actually, the mean $V$ and $I$ magnitudes from HST, for
instance, have grossly different errors, because the former are based
on a sufficient number of observations, but the latter only on a bare
minimum. Moreover, the $I$ magnitudes are measured against stronger
background effects. If a period-luminosity-color (PLC) relation is
used, as the author proposes, the necessarily large errors in $(V-I)$
or any other color are multiplied by the coefficient of the color term
of roughly 2.5 (depending on wavelength). A yet more fundamental objection
against the simple ansatz of a PLC relation has since been raised by
Saio \& Gautschy (1997) who find a strong variation of the slope
of the constant-period lines depending on the period. Kochanek offers
four different model solutions for the distance moduli, none of which
is fully self-consistent.  His solutions make the calibration in
eq.~(3) 0\fm 14 fainter on average. Yet we see no merit in these
solutions.

It remains the question of the metallicity dependence of the
Cepheid distances. Important effects have been proposed (e.g.~Sasselov
et al.~1997; Beaulieu \& Sasselov 1996; Sekiguchi \& Fukugita 1997).
On the other hand Madore \& Freedman (1991) and B\"ohm-Vitense (1997)
have given observational evidence against a metallicity dependence.
The most direct observational evidence comes from the metal-rich
Galactic and the metal-deficient LMC Cepheids; the former as calibrators
yield an LMC Cepheid distance in agreement with evidence that is
independent of Cepheids (Table 1). Also Di Benedetto (1997) has derived
Cepheid distances including the metal poor SMC, and has obtained fully 
consistent results {\sl independent} of wavelength. The question comes
now to rest with fully theoretical treatments of stellar evolution
coupled with pulsational theory indicating that the dependence
of $M_{\rm bol}$ on $\log P$ is surprisingly insensitive to metallicity
(Stothers 1988; Chiosi, Wood, \& Capitanio 1993; Sandage 1996c; 
Saio \& Gautschy 1997). Allowing for the metal dependence of the bolometric
correction introduces at a given $\log P$ a variation of $M_{\rm V}$
and of $<0\fm 1$ over the relevant metallicity range.

The {\sl slope} of the P-L relation of LMC is quite well determined
in different wavelengths (Madore \& Freedman 1991). Any remaining
uncertainties of the slope have a minute effect on the derived
distances. The period range of the Cepheids used for the calibrators
in Table 2 is sufficient to minimize selection effects (Sandage 1988);
in any case they can only lead to an underestimation of the
distances.

For these reasons it appears that any {\sl systematic} errors of the Cepheid
distances of the calibrators in Table 2 are below 0\fm 10.

\subsection{Virgo spiral galaxies}

The Virgo Cluster Catalogue (VCC; Binggeli, Sandage, \& Tammann 1985)
is the main base from which we selected the different samples of Virgo
spirals. The present investigation is restricted to Hubble types 2 to
8 (Sab to Sdm), because none of the calibrators is of type 1 (Sa), and
because late-type galaxies (type 9 and 10) have exceptionally uncertain
inclinations $i$.  The VCC contains 151 spirals of types 2 to 8 with
$v_0 < 2600$ km s$^{-1}$. Of these 98 (123) have $i>45^{\circ}
(30^{\circ})$, generally considered to be a necessary condition to
confine the inclination corrections of the observed line
width. Binggeli, Popescu, \& Tammann (1993) assigned memberships in
subgroups to many of the VCC galaxies. 34 \lq\lq normal\rq\rq ~spirals
with $i>45^{\circ}$ in the notation of Binggeli et al.~(1993) belong
to subgroup \lq\lq big A\rq\rq ~around M$\,$87, 15 are members of
subgroup \lq\lq B ($r=2.4^{\circ}$)\rq\rq , i.e.~they lie within
$2.4^{\circ}$ of M$\,$49. The corresponding numbers for galaxies with
$i>30^{\circ}$ are 48 and 19, respectively. For these galaxies
photometric as well as H$\,$I data are available from different
sources. We define these 34 \lq\lq inner\rq\rq ~spirals of A and the
15 \lq\lq inner\rq\rq ~spirals of B with $i>45^{\circ}$ as our \lq\lq
fiducial sample\rq\rq . The qualification \lq\lq normal\rq\rq ~is to
mean that they are not classified as \lq\lq peculiar\rq\rq ~(Binggeli
et al.~1985) and that they are not H$\,$I-truncated (Guhathakurta et
al.~1988).  The \lq\lq fiducial sample\rq\rq ~is identified in Table
3. The 5 H$\,$I-truncated and the 2 (5) peculiar spirals with
$i>45^{\circ} (30^{\circ})$ are listed next in Table 3.
Also listed are the 15 spirals which fall within the X-ray contours of
the cluster but outside of A and B.  All spirals of the \lq\lq
fiducial sample\rq\rq ~are considered to be members of the Virgo
cluster. This assignment will be tested below. Finally, Table 3
contains 41 galaxies which lie outside A and B and outside the X-ray
contours. Of these 15 lie outside the field of the VCC and were added
from other Virgo samples (Fouqu\'e et al.~1990; Kraan-Korteweg et al.~1988;
Pierce and Tully 1988). The projected positions of all the 103
non-peculiar spirals with $i>45^{\circ}$ is shown in Fig.~1.

\section{Observational Data}

\subsection{Calibrators}

The observational parameters of the calibrators and their sources 
are given in Table 2.
The columns have the following meaning:
\begin{enumerate}
 \item name in the NGC catalogue
 \item Hubble type as given by the RSA (Sandage \& Tammann 1987)
  in the nomenclature of the RC3 (de Vaucouleurs et al.~1991)
 \item total apparent $B$ magnitude corrected for internal and Galactic
  absorption as listed in the RC3 [based on the maps of Burstein \& Heiles
  (1984)]
 \item correction for internal absorption calculated from the Hubble type
  [column (2)] and $\log R_{25}$ following the RC3 scheme
 \item correction for Galactic absorption as listed in the RC3
 \item adopted true distance modulus 
 \item source for (6):
 \newcounter{zahl}
 \begin{list}{\arabic{zahl}}{\usecounter{zahl}}
 \item Madore \& Freedman (1991)\hfill \break $\dagger$ 
  B\"ohm-Vitense (1997) gives a Cepheid modulus of $(m-M)_{\rm M\,31}=24.67$ 
  if $(m-M)_{\rm LMC}=18.50$
 \item Silbermann et al.~(1996a)
 \item Silbermann et al.~(1996b)
 \item Sakai et al.~(1996)
 \item Freedman \& Madore (1994)
 \item Graham et al.~(1996)
 \item Tanvir et al.~(1995)
 \item Macri et al.~(1996)
 \item Ferrarese et al.~(1996)
 \item Saha et al.~(1996b)
 \item Sandage et al.~(1996)
 \item Kelson (1996)
 \item Hughes (1996)
\end{list}
 \item absolute $B$ magnitude calculated from (3) and (6)
 \item adopted error in the absolute magnitude [combined from the
  respective errors of the distance modulus as listed in the original source, 
  an assumed error of 0\fm 1 in the photometry and and assumed error of
  $0.2 (A^i + A^0)$]
 \item inclination calculated from $\log R_{\rm 25}$ as listed in the RC3
  following Fouqu\'e et al.~(1990);\hfill \break
  $^{\ast}$ adopted inclinations are $i=29^{\circ}$ for N$\,$4321 (mean from 
  Grosb\o l 1985, Warmels 1988, Arsenault et al. 1988, Pierce 1994)
   and $i=22^{\circ}$ for N$\,$5457 (mean from Rogstad \& Shostak 1972,
   Comte, Monnet, \& Rosado 1979, Bosma, Goss, \& Allen 1981, Pierce 1994)
 \item $\log$ H$\,$I line width read at the 20\% level, 
  corrected for inclination and turbulent motions and motions in $z$-direction
  (in most cases copied from Huchtmeier \& Richter 1988, 1989)
 \item $\log$ H$\,$I linewidth read at the 20\% level, corrected for
  inclination and turbulent motions and motions in $z$-direction, calculated 
  from $w_{20}$ given by LEDA. Inclinations for the edge-on corrections were
  taken from column (9) and the correction for turbulent motion and the
  motion in $z$-direction was calculated following Huchtmeier \& Richter 
  (1989).
 \item uncertainty in $\log$ H$\,$I linewidth composed of the assumed 
  uncertainties of the  measured linewidth $w_{20}$ and the inclination 
  of column (10), i.e.~10 km s$^{-1}$ and $5^{\circ}$, respectively.
\end{enumerate}

\subsection{Virgo spiral galaxies}

The data for the Virgo sample galaxies are listed in Table 3. The columns
have the following meaning:

\begin{enumerate} 
\item Name as listed in one of the following catalogs: NGC (N), IC (I),
UGC (U), VCC (V), ZWG (Z)
\item Virgo subgroup membership as assigned by Binggeli et al.~(1993);
Y means that the galaxy does not lie in VCC field but is included in
the samples of other authors; spirals with code Z are not assigned a
subgroup membership; X means member of the X-ray sample (the projected
position is within the defining X-ray isophote; cf.~Sect.~5.3)
\item Hubble type as listed in the VCC on the scheme of the RC3
\item total apparent $B_{\rm T}$ magnitude from RC3, corrected for
internal and Galactic absorption as given in columns (6) and (8)
\item uncertainty in $B_{\rm T}^{0,i}$ composed of the
assumed errors of the photometry and of the internal and Galactic
absorption corrections, i.~e.~0\fm 1 and $0.2 (A^i+A^0)$,
respectively; for the fit of the TF relation we assume in addition an
uncertainty of 0\fm 2 to make some allowance for the depth effect of
the cluster.
\item correction for internal absorption calculated for the
Hubble type listed in column (3) and $\log R_{25}$ as given in the RC3
following the RC3 scheme
\item correction for internal absorption as given in the RSA
\item $A^0$ correction for Galactic absorption as given in the RC3
[based on the maps of Burstein \& Heiles (1984)]
\item inclination calculated from $\log R_{\rm 25}$ as listed in the RC3
following the recipe of Fouqu\'e et al.~(1990)
\item $\log$ H$\,$I line width read at the 20\% level, corrected for
inclination and turbulent motions and motions in $z$-direction, taken from
Huchtmeier and Richter (1989). An asterisk denotes the cases where $\log w$
had to be calculated from $w_{20}$ listed in the RC3. Two asterisks indicate
those cases where $\log w$ was computed from $w_{20}$ given by Hoffman et
al.~(1987, 1989). Cases for which no H$\,$I data were available are marked
with a dagger.
\item $\log$ H$\,$I line width read at the 20\% level, corrected for
inclination and turbulent motions and motions in $z$-direction,
calculated from $w_{20}$ given by LEDA. Inclinations were taken from
column (9) and the correction for turbulent motion and the motion in
$z$-direction was calculated following Huchtmeier \& Richer (1989).
\item uncertainty in $\log$ H$\,$I linewidth composed of the assumed 
uncertainties of the  measured linewidth $w_{20}$ and the inclination 
of column (9), i.e.~10 km s$^{-1}$ and $5^{\circ}$, respectively.
\end{enumerate}

\section{The calibrated Tully-Fisher relation}

A first solution of the calibration of the Tully-Fisher relation was
obtained by fitting a straight line to the absolute magnitudes (as the
dependent variable) and line widths (as the independent variable) of the
calibrators listed in Table 2, columns (8) and (12). Each calibrator
was weighted with its estimated error of $M_{\rm B}$ (column 9 of
Table 2). The result of this free fit is:
\begin{equation} 
M_{\rm B}= (-6.32\pm 0.08) \log w - (4.05\pm 0.19) \ \ \ \ \sigma_M=0\fm 43. 
\end{equation} 
A second solution is presented taking the errors in magnitude
{\sl and} line width into account. This is achieved by dropping
slanted lines with slope $\epsilon_y/\epsilon_x$ onto a regression
line and demanding a least squares minimalization (cf.~Seares 1944).
For this the BCES (Bivariate Correlated measurement Errors and
intrinsic Scatter) algorithm of Akritas, Bershady \& Bird (1996) is
used. The adopted errors on both axes are taken from columns (9) and
(13) of Table 2 and from columns (5) and (12) in Table 3.  Since the
slope of the TF relation is determined much better by including the 49
members of the \lq\lq fiducial Virgo sample\rq\rq ~described in
Sect.~2.2 we combine the sample of the calibrators with the Virgo
sample assuming different distance moduli and search for that solution
which gives a minimum value of the merit function $\chi^2$. 

We obtain the following relation which is illustrated in Fig.~2:
\begin{equation} 
M_{\rm B}= (-6.97\pm 0.02) \log w - (2.35\pm 0.05).
\end{equation}
The slope of eq.~(3) is somewhat steeper than the slope of eq.~(2), a 
consequence of now including the errors in $\log w$. 

We adopt eq.~(3) as our reference and discuss the influence of
different input parameters in Sect.~6. The effect of the remaining
uncertainty of the slope is discussed in Sect.~10.

It is of some theoretical interest to determine the intrinsic scatter
of the TF relation. Taking the observed magnitude scatter about
eq.~(3) of $\sigma_{\rm M}=0\fm 44$ from the calibrator sample and
subtracting the typical error in $M$ of the calibrators, i.e.~0\fm 25,
and the error introduced by the mean adopted errors in line width,
which translate to $\sigma_{{\rm M}(w)}=0\fm 28$, gives an intrinsic
scatter of $\sigma_{\rm M}=0\fm 23$. In the presence of unavoidable
errors of the input parameters this value is, of course, meaningless
for practical applications.

\section{The Virgo cluster distance}

In this section the TF distances to different Virgo samples are discussed.

\subsection{A first approximation}

The direct TF relation allowing only for errors in magnitude [Table 3, 
columns (4) and (5)] gives for the  \lq\lq fiducial sample\rq\rq ~defined 
in Sect.~2.2
\begin{equation} 
m_{\rm B}= (-6.60\pm 0.03) \log w + (28.31\pm 0.06) \ \ \ \ \sigma_M=0\fm 57. 
\end{equation}
and combined with eq.~(2) gives a distance modulus of $(m-M)= 31\fm 62\pm 
0\fm 08$. We prefer, however, the solution which takes into account the 
errors in magnitude {\sl and} line width. In this case a $\chi^2$ method 
(cf.~Sect.~4) gives
\begin{equation} 
(m-M)= 31\fm 50\pm 0\fm 09 .
\end{equation}
The data were weighted with their individual errors in magnitude and
line width.  This value is further discussed in Sects.~6.5 and 8. The 
finally adopted value and its error are given in Sect.~10.

If the spirals in Fig.~3 are subdivided into subclusters A and B
a significant distance difference of $\Delta (m-M)=0\fm 46\pm 0\fm 18$
emerges (cf.~Schr\"oder 1996). This raises several problems. The
distance difference is small enough and the distance overlap between
individual galaxies so pronounced (cf.~Figs.~11 and 12 below) that the
two subgroups must be gravitationally bound as is also shown by their
common X-ray contour (cf.~Fig.~4 below). Moreover, the distance
difference is {\sl not} confirmed by other relative distance
indicators. $D_{\rm n}-\sigma$ data (Faber et al.~1989) for 11
galaxies in A in B suggest A to be {\sl more} distant than B by $0\fm
05\pm 0\fm 36$; two blue SNe$\,$Ia and one in B give $\Delta
(m-M)=+0\fm 55\pm \sim 0\fm 3$ again in the sense of A being more
distant. The apparent magnitude of NGC\,4472 (M\,49), the brightest
galaxy in B, is already brighter than that of NGC\,4486 (M\,87) in
subgroup A which has all the properties expected of a brightest
cluster galaxy; if B is assumed to be more distant than A the {\sl
absolute} magnitude of NGC\,4472 becomes excessively bright as
compared to NGC\,4486.  Not attempting here to understand the complex
spatial structure of the cluster, we must adopt for practical reasons
a mean distance of the spirals in subgroup A {\sl and} B because all
previous workers in the field have done so. In particular, the Hubble
diagram with cluster distances relative to the Virgo cluster (Fig.~14
below) is based on the {\sl entire} Virgo sample.

The \lq\lq fiducial sample\rq\rq ~contains only galaxies with
inclinations $>45^{\circ}$ to avoid possibly large errors in the
calculation of true line widths from the observed line widths. If an
inclination limit of $i>30^{\circ}$ is admitted instead of
$i>45^{\circ}$, the sample is increased from 49 to 67 \lq\lq normal\rq\rq
~members in subgroups A and B. This increased sample gives instead of
eq.~(5)
\begin{equation} 
(m-M)=31\fm 48 \pm 0\fm 07. 
\end{equation}
The TF slope of the combined samples (Virgo [$i>30^{\circ}$] and
calibrators) is $-7.15$. The effect of the less inclined galaxies on the 
distance modulus is surprisingly small. 

The Virgo cluster modulus is further discussed in the following
subsections and in Sect.~6.

\subsection{The influence of H$\,$I-truncated and peculiar galaxies on the TF 
distance}

There are five galaxies in subgroup A which have an unusually small H$\,$I
disk (Guhathakurta et al.~1988). These galaxies were excluded from the \lq\lq
fiducial sample\rq\rq . If the H$\,$I-stripped objects listed in
Table 3 were included into the calculation of the TF distance of the fiducial
sample one would obtain $(m-M)=31\fm 43\pm 0\fm 09$. To show the effect 
even more
drastically: the mean distance of the H$\,$I-stripped galaxies {\sl alone} 
with the calibration of eq.~(3) is $(m-M)=30\fm 81\pm 0\fm 24$.
Figure 3 shows that the H$\,$I-truncated galaxies
are $\sim 0\fm 8$ {\sl too luminous for their line width}. 
We conclude that it is not appropriate to use H$\,$I-truncated galaxies in
the TF relation. Distances to samples with H$\,$I truncated galaxies are
underestimated.

The same conclusion holds for the two galaxies of the fiducial sample
which are classified as peculiar or interacting from their
morphological appearance. With a mean $(m-M) = 31\fm 14\pm 0\fm 21$ 
they also tend to give too small distances. They were therefore excluded 
from our analysis.

\subsection{The distance of the Virgo X-ray sample}

The diffuse X-ray emission of the hot gas in the Virgo cluster has
been measured by ROSAT (B\"ohringer et al.~1994). We use the X-ray
boundary, i.e.~the isophote with 0.444 counts s$^{-1}$ arcmin$^{-1}$
as an {\sl independent} mean to define the cluster population
(Fig.~4). There are 42 galaxies with $i>45^{\circ}$ within the
defining X-ray contours (H$\,$I-truncated and peculiar objects
omitted). The fact that the south-east side of the X-ray contour is
disturbed by the Galactic North Polar Spur has only negligible effect
on the sample selection.

Combining both the calibrator and the X-ray samples and taking errors
both in magnitudes and $\log w$ into account gives a TF slope of $-6.55$,
which is slightly but insignificantly shallower than that found in eq.~(3). 
The distance modulus
of the X-ray sample corresponding to this solution is
\begin{equation}
(m-M)= 32\fm 05\pm 0\fm 11, \ \sigma_{(m-M)}=0\fm 70. 
\end{equation} 
The distance is 0\fm 55 larger than the distance of the optically
selected \lq\lq fiducial sample\rq\rq . The difference is partly due
to the seven galaxies in the X-ray sample which are assigned to the W
and W' cloud (lying in or near the lower right \lq\lq peninsula\rq\rq
~of the defining X-ray isophote in Fig.~4). The latter have a mean
recession velocity of $v_{\rm LG}=1645$ km s$^{-1}$ and give a TF
distance modulus of $(m-M)=32\fm 58 \pm 0\fm 20$. The X-ray sample 
without these seven galaxies yields 
\begin{equation}
(m-M)=32\fm 00\pm 0\fm 11 \ \ \ \ \ \sigma_{(m-M)}=0\fm 68.
\end{equation}

Even after exclusion of the seven galaxies the distance modulus is
relatively high. The reason is the shallower slope of the TF
relation of the X-ray sample as compared to the fiducial sample.
Shallower slopes always tend to increase the distance.

\subsection{The distances of Virgo galaxies outside the \lq\lq fiducial 
sample\rq\rq ~and outside the X-ray sample}

If the adopted TF relation (eq.~3) is applied to the 38 galaxies of
Table 3 which are not members of the fiducial sample or the X-ray
sample, one obtains a large mean distance modulus of $ (m-M)=31\fm 85$
with a large dispersion of $\sigma_{(m-M)}=1\fm 05$. Indeed, Binggeli
et al.~(1993) assigned some of these outlying members to the M, W, W'
or S clouds, which appear to be in the background for morphological
and other reasons.

The individual TF distances of the outlying galaxies are plotted versus 
recession velocity in Fig.~5. The qualitative difference to Fig.~6 below is
obvious. Some nearby and several distant galaxies are probably not
cluster members. They seem to partake in the (nearly) free expansion
field.

\section{The influence of different data sources on the TF distance of
the Virgo cluster}

The influence of different input parameters on the distance of the
Virgo cluster is now investigated. 

\subsection{Using alternative $B$ magnitudes}

The distance of the Virgo cluster does not change when
magnitudes from different sources are used. Replacing the $B_{\rm T}$
magnitudes in Table 3, column 4 by those of Schr\"oder (1996), but
keeping the RC3 corrections, increases the distance modulus by only
0\fm 04. Using instead the magnitudes of of Yasuda, Fukugita, \&
Okamura (YFO 1997), again with the RC3 corrections, gives a very similar
modulus of $(m-M)=31\fm 49$ for the fiducial sample. 

\subsection{Using an alternative set of line widths}

Huchtmeier \& Richter (1988, 1989) have published H$\,$I data for nearby
galaxies and galaxies in the Virgo cluster.  Their line widths can be
used as an alternative data set to the $w_{20}$ data of the Lyon
Extragalactic Data Base (LEDA). The two sets of $w_{20}$ line widths 
usually are in good agreement, although there 
is an average systematic offset of +6 km s$^{-1}$, the Huchtmeier \&
Richter values being larger. Using the fully reduced Huchtmeier \&
Richter data for the calibrators as well as for the Virgo spirals
gives the same distance within 0\fm 09 as eq.~(5). However, the
dispersion of the TF relations for both the calibrators and the Virgo
galaxies is slightly enlarged (Table 4, solution 2).

\subsection{Using an alternative scheme for internal absorption}

So far the corrections for internal absorption were taken from the RC3.
Alternative precepts to determine the internal-absorption corrections
are given in the RSA (San\-dage \& Tammann 1987). The main difference is
that the RC3 corrects the magnitudes to face-on orientation, while
the RSA allows for the {\sl total} internal absorption. In addition
the RSA corrections have a somewhat stronger dependence on Hubble
type. As seen from Table 4 for solutions 5 and 6, the internal absorption
corrections of the RSA lead to a slightly steeper slope of the
TF relation and to an insignificantly larger scatter. Also the
Virgo modulus is $\sim 0\fm 17$ larger. However, the individual
TF distances are equally consistent as those with the RC3 correction,
i.$\,$e.~they do not correlate significantly with Hubble type,
line-width, nor inclination (cf.~Sect.~7). The modulus
difference between subclusters A and B remains at 0\fm 46. 

It has often been proposed to use infrared magnitudes for the TF
relation to circumvent the relatively large intrinsic-absorption
corrections in $B$. However, the absorption in the $I$-band is still
much larger than a $1/\lambda$-law would suggest (Schr\"oder 1996;
Giovanelli et al.~1994). Moreover, the modest advantage of the smaller
absorption is offset by the steeper slope of the $I$-band TF relation
(for a discussion cf.~Schr\"oder 1996).

\subsection{Corrections for Galactic absorption}

Since the Galactic latitude of the Virgo cluster is $60^{\circ} < b <
80^{\circ}$ the influence of Galactic absorption is small or negligible.
Maps of the H$\,$I column density indicate that the Galactic absorption is
not zero for all Virgo cluster galaxies but varies between 0\fm 00
and 0\fm 16 (Burstein \& Heiles, 1984). These corrections, as listed 
in the RC3 and whose mean value is only 0\fm 05, have been applied to all 
galaxies in Table 2 and 3.

If, instead, one assumes $A^0=0$ for the Virgo galaxies, the slope of the
TF relation is not changed within the error
limits; the dispersion becomes insignificantly smaller (Table 4, solution 8).

\subsection{The effect of different input parameters}

The solutions 1-8 in Table 4 (column 8) give an overview how the Virgo cluster
modulus varies as different sources of the magnitudes, line widths,
and internal absorptions are used. Also shown is the effect
of increasing the sample of spirals with $i>45^{\circ}$ by those
with  $30^{\circ}<i<45^{\circ}$ and by replacing the optically selected
fiducial sample by the members within the X-ray contour. Finally
in solution 8 the small effect of assuming zero
Galactic absorption for the cluster members is shown.

The resulting Virgo cluster moduli in column 8 span a range
of 31\fm 48 to 32\fm 00. There is no objective way to
discriminate between the different solutions. We therefore
take a straight mean over the solutions 1-7 and obtain
\begin{equation}
(m-M)_{\rm Virgo}=31\fm 65 \pm 0\fm 07 .
\end{equation}

\section{Testing the internal consistency of the results}

Several tests are performed to check the internal consistency of the
derived TF distances. Without loss of generality we consider here
only the TF distances of solution 1 in Table 4.

The individual distance moduli are shown in Fig.~6 as a function of
redshift.  There is no significant correlation between the TF distance
and the recession velocity.  From Fig.~6 one could argue that the five
galaxies with redshifts larger than the gap at $v_{\rm LG} \sim 1700$
km s$^{-1}$ do not belong to the cluster. However, the gap is a fluke
of small sample statistics, because it does not appear in the velocity
distribution of {\sl all} Virgo members (Binggeli et al.~1993).
Moreover the high-velocity galaxies have closely the same mean TF
distance as the rest of the fiducial sample. There is no objective
argument to exclude any galaxy in Fig.~6 from cluster membership. The
only statistically significant effect is the dichotomy between
subclusters A and B as discussed in Sect.~5.1.

Figure 7 shows the individual TF distances plotted against inclination.
Any systematic error of the inclinations affects both the corrected
line widths and the corrections for internal absorption and translates
into a dependency of the TF distances on inclination. However, no such
trend exists, not even for the spirals with $30^{\circ}<i<45^{\circ}$.

Similarily we do not find a dependence of the TF distances on the
Hubble type (Fig.~8). This would have been the case if the corrections for
internal absorptions, which are a function of the Hubble type, were
systematically wrong or if the TF relation was significantly different for
different Hubble types. (The question whether TF distances depend on
Hubble type is largely determined by the internal-absorption correction
used.)

As a next test we show a plot of the individual TF distances as a function
of line width (Fig.~9). The obtained distances do not depend on the line width
which must be the case if the slope of the TF relation has been determined
correctly.

In Fig.~10 there is a trend of galaxies becoming fainter with
increasing TF distances. This can most naturally be explained by the
cluster depth effect.  If the galaxies are randomly distributed in the
cluster as to absolute magnitude $M_{\rm B}$, then the nearer galaxies
must be apparently brighter.  A least square fit through the data
gives $(m-M)= (0.12\pm 0.06) m_{\rm B} + 29\fm 95$. This implies that
the cluster has an effective depth of $\pm 0\fm 4$ as outlined by the
spirals of the five-magnitude interval $10^{\rm m}<m_{\rm B} <15^{\rm
m}$.

The eight galaxies fainter than $14^{\rm m}$, which all seem to lie on
the far side of groups A and B respectively, do not show an unusual
behaviour in the correlation of the parameters in Figs.~6-9. Two of
the eight galaxies, i.e.~NGC$\,$4353 and IC$\,$3298, have
particulary large TF distance moduli of $(m-M)\approx 32\fm 6$. In
spite of this NGC$\,$4353 must be assigned to the cluster with a
recession velocity of only $v_{\rm LG}=982$ km s$^{-1}$. It is
therefore arbitrary to exclude any galaxy of the fiducial sample from
cluster membership only on the basis of its large TF
distance. Therefore, there is also no rational reason to exclude IC
3298 ($v_{\rm LG}=2355$ km s$^{-1}$) from the cluster.

\section{Systematic differences between calibrators and Virgo spirals}

The Virgo cluster distance depends on the {\sl assumption} that field
and cluster galaxies obey the same TF relation.  This assumption can
now be tested with the complete {\sl UBVRI} photometry of Schr\"oder
(1996). It turns out that the cluster galaxies are redder in $(B-I)$
(corrected for internal and Galactic reddening) at a given line width
than the 18 calibrators which are predominantly field galaxies.  This
means that the TF relation cannot be the same for the two sets of
galaxies {\sl in all wavelengths}. The cluster galaxies are also more
H$\,$I-deficient than the calibrators. The hydrogen deficiency $D_{\rm
H\,I}$ is here defined following YFO (1997), whereby the color
residuals $\Delta (B-I)$ at fixed line width correlate well with
$D_{\rm H\,I}$. The dependency of the individual TF distances on
$\Delta (B-I)$ and $D_{\rm H\ I}$ are shown by Schr\"oder \& Tammann
(1996) for five different passbands. Tammann \& Federspiel (1997)
have argued that the best Virgo distance is obtained by averaging over
all color and $D_{\rm H\,I}$-corrected {\sl UBVRI} moduli. The result
is that the straight TF modulus in $B$ is $0\fm 07\pm 0\fm 02$ {\sl
larger} than the mean of all corrected moduli.  We apply this
correction to eq.~(9) to obtain the finally adopted Virgo modulus of
\begin{equation} 
(m-M)^0_{\rm Virgo}= 31\fm 58\pm 0\fm 07 
\end{equation}
where the error is taken from eq.~(9).

It may be noted in passing that the TF modulus in $V$ seems least sensitive
to the cluster effect and that the {\sl uncorrected} TF modulus in
$I$ is 0\fm 09 too small (Schr\"oder \& Tammann 1996).

\section{The three-dimensional structure of the Virgo cluster from TF data}

Individual Tully-Fisher distances are reliable enough that they can be
used to cautiously explore the three-dimen\-sional structure of the Virgo 
cluster. Figure 11 shows a visualisation of the spatial distribution
of the members of subgroups A and B as well as of the calibrators that
belong to the Virgo cluster. All distances given in this section are
based on Table 4, solution 1. The $x$-$y$ plane is perpendicular to the
line-of-sight from the Sun to M$\,$87, i.e.~it is a tangent plane to the
celestial sphere. The spherical coordinates were transformed into linear 
coordinates at the distance of the individual galaxies. Since the distance 
moduli obtained from the TF relation are a logarithmic measure, the 
$z$ axis is logarithmic ($\log$ distance in Mpc). A typical error of an
individual distance of the order of the scatter of the TF relation
[0\fm 6 in $(m-M)$] corresponds to an uncertainty of 0.1 in $z$ on the line
of sight to M$\,$87.  

The calibrators among the Virgo galaxies and NGC 4496A are shown as
crosses at their respective {\sl Cepheid} distance. Three of them
(N4321, N4496A, and N4536) are among the dozen closest Virgo galaxies,
whereas N4639 is on the far side of subcluster A. These four galaxies
with known Cepheids provide a check on the accuracy of about 0\fm 45 
of the individual TF distances (cf.~Table 5). 

From the apparent angular size and the 
depths calculated from the TF distances (deconvolved with the
intrinsic dispersion of about 0\fm 40 of the TF relation) we derive
a linear diameter-to-depth ratio for subcluster A of about 1:1.5.

Figure 12 emphasizes that subcluster B on average lies at a greater
distance than subcluster A ($0\fm 46\pm 0\fm 18$ in the distance modulus). B
appears rather elongated in depth although it is quite compact in the
projection on the celestial sphere. The diameter-to-depth ratio for B is
about 1:5. The ratios may be somewhat smaller if we have
underestimated the distance errors.

\section{Discussion}

The error of the adopted TF modulus of the Virgo cluster in eq.~(10) 
reflects only the random error. Systematic error sources are listed
in Table 6.

Remarks to Table 6 are as follows. The zeropoint error of the P-L
relation has been discussed in Sect.~2.1. Forcing extreme slopes of
$-7.50$ and $-6.50$ on the TF relation of the fiducial sample and the
calibrating galaxies changes the distance modulus by $\pm 0\fm 10$.
Different precepts of weighting the data of the calibrators and Virgo
spirals can influence the modulus by 0\fm 08. Giving higher
weight to the X-ray sample or omitting it altogether could change the
adopted distance modulus by 0\fm 1 at most. Taking the input $B$
magnitudes and/or line widths $w$ from different sources has
essentially no effect on the Virgo distance (Sects.~6.1.~and 6.2.);
but different precepts of the internal-absorption correction affect
the modulus by $\pm 0\fm 10$ (Table 4, solutions 1-3 versus 5-7).
Different properties of the calibrators and the cluster galaxies have
been accounted for by excluding the truncated cluster galaxies and by
applying a mean correction for differences in $(B-I)$ and $D_{\rm
H\,I}$ (Sect.~8); the remaining maximum error is estimated to be
$\pm 0\fm 05$. The calibrators are systematically {\sl less} inclined
($\langle i \rangle = 55^{\circ}$) than the cluster galaxies ($\langle
i \rangle = 72^{\circ}$). Consequently the mean RC3 correction for
internal absorption is $\langle A^i \rangle = 0\fm 34$ for the former
and $\langle A^i \rangle = 0\fm 66$ for the latter. Yet the two
samples become more similar if the fiducial sample is increased by the
cluster galaxies with $30^{\circ}<i<45^{\circ}$. In that case the
cluster sample has $\langle i \rangle = 62^{\circ}$ and $\langle A^i
\rangle = 0\fm 45$ without changing the cluster modulus by more than
0\fm 03 (Table 4, solutions 3 and 7). Moreover the individual TF
distances are independent of inclination (Fig.~7), which would not be
the case if the internal absorption corrections were inadequate.  The
conclusion is that the internal absorption corrections introduce a
systematic error of not more than 0\fm 05.

Summing the errors in quadrature in Table 6 gives an external error of
$\pm 0\fm 23$. Adding this in quadrature to the internal error
in eq.~(10) gives the final distance modulus of the Virgo
cluster (A and B) to be 
\begin{equation} 
(m-M)_{\rm Virgo}^0 = 31\fm 58 \pm 0\fm 24 \ ({\rm external\ error}) 
\end{equation}
or $r_{\rm Virgo}= 20.7 \pm 2.4 {\rm \ Mpc}$. 

The decisive element of the present distance determination of the
Virgo cluster is that a {\sl complete} sample has been used.  It is
fortunate that low-luminosity spirals {\sl do not exist}, and that the
Virgo Cluster Catalog (VCC; Binggeli et al.~1985) goes much fainter
than the faintest spirals. The faintest galaxies in that catalog are
of morphological types Im and dE, well fainter than the absolute
magnitude limit of the TF calibrators of Table 2 and Fig.~2. The
sample studied here is therefore indeed {\sl complete} for
spirals. The importance of completeness is illustrated in Fig.~13,
where the fiducial sample is cut by different magnitude cutoffs.  The
brighter the cutoff is the smaller is the distance modulus even for
{\sl complete} flux-limited samples (cf.~Teerikorpi 1987).

Giovanelli et al.~(1997a, b) have given TF data for inclined spirals
in 24 clusters. There are 23 galaxies per cluster on average which
reach $\sim 3\fm 8$ into the luminosity function on average. This
would be barely sufficient to derive an unbiased distance from a
complete sample (Fig.~13). However, the cluster samples are
admittably {\sl incomplete} at all levels but particularly at the
fainter magnitudes. The authors derive excellent {\sl relative}
cluster distances as seen in Fig.~14 below, due to a very careful
modelling of the cluster incompleteness. They have also attempted
to obtain an absolute calibration by fitting the cluster data to
the local distance-limited calibrators and suggest $H_0=69\pm 5$
(Giovanelli et al.~1997c). Clearly this latter step is much more
sensitive to the bias model than the relative distances. If the
authors had used the Virgo cluster distance of eq.~(11) they would have
obtained from their cluster data $H_0=57$ as below in eq.~(15).
Vice versa their value of $H_0=69$ would require 
$(m-M)_{\rm Virgo}=31\fm 17$ which we deem to to be highly improbable.

In a recent application of the TF method to the Virgo cluster YFO
(1997) have analyzed a less restrictive sample of 108 inclined Virgo
spirals and irregulars. Their mean distance is $(m-M)=31\fm 81$ with a
large scatter of $\sigma_{(m-M)}=0\fm 93$. Restricting the sample to
the 63 galaxies with $\delta > 10.5^{\circ}$ -- which roughly
corresponds to subcluster A -- yields $(m-M)=31\fm 62$ with
$\sigma_{(m-M)}=1\fm 00$. These moduli are somewhat larger than those
of the fiducial sample and subcluster A, respectively, from above.
The scatter of the latter solutions is significantly less. Moreover
YFO did not correct the line widths for turbulence, resulting in a
quite steep TF relation which leads to a significant dependence of the
individual distances on the corrected line widths. The extremely low
cluster modulus of $(m-M)=31\fm 09$ finally adopted by YFO for the
M$\,$87 subcluster was achieved by confining their sample to only
galaxies with $\delta > 10.5^{\circ}$ and by excluding 21 spirals with
$(m-M)>32\fm 0$. We take exception to these {\sl artificial} cuts
which prevent any objective distance determination of the Virgo
cluster. There is no objective reason to exclude the more distant
galaxies which clearly are members. They not only perfectly fit under
the distance distribution curve (cf.~Fig.~12) but also are
indistinguishable from other cluster members on the basis of both
their positions within the optical and X-ray contour of the cluster in
the sky and also their velocities.

Shanks (1997), using both Cepheid and SN$\,$Ia distances as calibrators
of the TF relation, has obtained an internally consistent
zeropoint correction of $+0\fm 46\pm 0\fm 11$ to the TF distance
of the Virgo cluster by Pierce \& Tully (1988), obtaining now
$(m-M)_{\rm Virgo}=31\fm 43\pm 0\fm 20$. Since the Pierce \& Tully sample
is {\sl incomplete} even this corrected distance must suffer
the Teerikorpi incompleteness bias and be systematically
too low (Fig.~13).

The Virgo modulus in eq.~(11) is identical within the errors
with the result of five independent methods,  including
Cepheids, SNe$\,$Ia, globular clusters, novae, and the $D_{\rm n}-\sigma$
method, giving $31\fm 66 \pm 0\fm 08$ (Sandage \& Tammann 1997).

\section{Conclusion}

\subsection{$H_0$ from the observed, infall-corrected velocity}

The Virgo cluster has been used frequently as a cosmic milestone to
derive the Hubble constant from its velocity and its distance. The
observed mean {\sl heliocentric} velocity of the cluster is
$v=1050\pm35$ km s$^{-1}$ (Binggeli et al.~1993) or $v_{\rm LG}=922$
km s$^{-1}$. This latter value must be transformed to the cosmic
recession velocity that takes the Virgocentric infall of the Local
Group into account. For the Virgocentric infall we adopt the value of
$v_{\rm infall}=220\pm 50$ km s$^{-1}$ (Tammann \& Sandage 1985),
which is in excellent agreement with $v_{\rm infall}=224\pm 90$ km
s$^{-1}$ (Bureau et al.~1996) and in statistical agreement with
$v_{\rm infall}=275\pm 90$ km s$^{-1}$ of Hamuy et al.~(1996); it is
further supported in Sect.~11.2.  Thus the infall-corrected
recession velocity of the Virgo cluster is $1142\pm 61$ km
s$^{-1}$. This combined with the adopted distance of the Virgo cluster
(eq.~11) gives a local Hubble constant of
\begin{equation} 
H_0=55\pm 7\ {\rm km\ s^{-1}\ Mpc^{-1}}. 
\end{equation}

Although the just described route to $H_0$ is quite common, it is
surpassed by the calibration of the Hubble diagram of clusters
addressed in the next section, which ties the Virgo cluster to
distant clusters and leads to a value of the Hubble
constant that is valid over a much larger volume. 

As an exercise we consider also the effect on $H_0$ if the
subclusters A and B are reduced separately. The data are set out
in Table 7. The mean velocity of the two aggregates corrected for an
infall velocity of 220 km s$^{-1}$ is taken from Binggeli et al.~(1993).
The distance moduli are calculated from solutions 1-3 and 5-7
in Table 4, but now separately for A and B. Here the X-ray sample
is, by necessity, not considered. The weighted mean of $H_0=56\pm 3$
reflects only the internal error (cf.~Sect.~10).

\subsection{$H_0$ from the Hubble diagram of galaxy clusters}

Cluster distances relative to the Virgo cluster are available for 17
clusters from various methods such as the Tully-Fisher (TF) method, the
$D_{\rm n}-\sigma$ relation, and first-ranked cluster galaxies (for a
compilation see Jerjen \& Tammann 1993). In addition, high weight
{\sl relative} TF distances are available for 24 clusters from
Giovanelli (1996). The latter list does not include the
Virgo cluster, but since eight clusters are in common, the two lists
can be merged with a mean error of only 0.05 mag. The double cluster
A$\,$2634/66 is not used here because Giovanelli (1996) only
gives a distance for A$\,$2634. The resulting Hubble diagram (Fig.~14)
contains 31 clusters with distances relative to the Virgo cluster
(cf.~also Table 1 of Tammann \& Federspiel 1997).

Clusters with $v_0 < 3000\,$ km s$^{-1}$ are corrected for a
Virgocentric infall model with a local infall velocity of $220\,$ km
s$^{-1}$. More distant clusters do not partake of the local motion
with respect to the CMB. They are therefore corrected for a CMB vector
of $630\,$ km s$^{-1}$. The dividing limit of $3000\,$ km
s$^{-1}$ is derived elsewhere (cf.~Jerjen \& Tammann 1993; Federspiel,
Sandage, \& Tammann 1994, Figs.~17-19; Giovanelli 1996); the exact choice
has no effect on the following conclusions.

The ridge line in Fig.~14 is represented by
\begin{equation}
\log v^{\rm CMB}= 0.2 \Bigl [(m-M)-(m-M)_{\rm Virgo}\Bigr ] 
                 +(3.070\pm 0.011).
\end{equation}

\noindent The slope of 0.2 is forced, the strongly deviating Eridanus
cluster is omitted.

Simple transformation of eq.~(13) gives
\begin{equation}
\log H_0 =\log v^{\rm CMB}-\log r_{\rm Mpc}
         =-0.2 (m-M)_{\rm Virgo} + (8.070\pm 0.011)
\end{equation}
Inserting the distance modulus of the Virgo cluster from eq.~(11)
into eq.~(14) gives the value of $H_0$ for distances as large as 
$11\,000\,$km s$^{-1}$:
\begin{equation}
H_0=57\pm 7\ {\rm km\ s^{-1}\ Mpc^{-1}}. 
\end{equation}

Note that reading eq.~(13) at {\sl zero relative distance} also
predicts the velocity of the Virgo cluster itself in the CMB frame
$v_{\rm Virgo}^{\rm CMB}=1175\pm 30\,$km s$^{-1}$. This is the
velocity one would observe in the absence of all local peculiar or
streaming velocities. Comparing this value with the actually observed
cluster velocity of $v_{\rm LG}=922\pm 35\,$km s$^{-1}$, one obtains a
Virgocentric infall velocity of the Local Group of $v_{\rm
infall}=253\pm 46\,$km s$^{-1}$ (cf.~Jerjen \& Tammann 1993). We take
this as a confirmation of $v_{\rm infall}=220\pm 50$ km s$^{-1}$
adopted in Sect.~11.1.

{\sl It must be stressed that the determination of $H_0$ from
eq.~(14) depends only on the quality of the Virgo cluster distance and
on the relative distances to other clusters, but
it is totally independent of any observed or inferred velocity of that
cluster} (Sandage \& Tammann 1996).

\subsection{Comparison with external values of $H_0$}

Several authors have suggested to derive $H_0$ at the position of
the Coma cluster (e.g.~Baum et al.~1995, 1996; Thomsen et al.~1997;
Hjorth \& Tanvir 1997). The result is typically $H_0\approx
70$. However, the experiment with only a single cluster is much more
difficult than the route via the Virgo cluster tied to the expansion
field of {\sl many} clusters (Fig.~14). The reason is, first, that
comparable objects in Coma are almost four magnitudes fainter than in
Virgo, and, secondly, that the velocity of Coma in the CMB frame is
quite poorly known. The observed mean velocity is $v_{\rm LG}=6912$ km
s$^{-1}$ (Zabludoff et al.~1993).  It is generally {\sl assumed} that
the cluster does not take part of the local motion relative to the CMB
frame. In that case $v_{\rm Coma}^{\rm CMB}=7194$ km
s$^{-1}$. However, clusters have random motions (in the radial
direction) of typically 450 km s$^{-1}$ (Jerjen \& Tammann 1993), and
Coma deviates indeed strongly from the Hubble line in Fig.~14. If a
modulus difference between Coma and Virgo of $\Delta (m-M)_{\rm
Coma-Virgo}=3\fm 72$ is adopted (Dekel 1996) eq.~(13) gives $v_{\rm
Coma}^{\rm CMB}=6500$ km s$^{-1}$.  Hence the velocity uncertainty
alone affects the result of $H_0$ by $\sim 10\%$.

The most direct way to derive the Hubble constant is through the
Hubble diagram of supernovae of type I$\,$a (SNe I$\,$a) which is by
now calibrated by seven SNe$\,$Ia with known Cepheid distances.
The result of $H_0=58\pm 8$ (external error)(Saha et al.~1997)
is in agreement with the present result. The only interdependence
of these determinations of $H_0$ is that two Cepheid distances used
to calibrate the SNe$\,$Ia are also used in Table 2 to calibrate the
TF relation.

Luminosity class distances and TF distances of field galaxies give
equally consistent values of the Hubble constant, after
correction of the Malmquist bias, of $H_0=55\pm 5$ (Sandage \& Tammann
1975; Sandage 1996a,b and references therein; Theureau et al.~1997).

The conclusion is that the three independent routes through (1) the
Virgo cluster, (2) SNe$\,$Ia, and (3) field galaxies all give $H_0=55\pm 5$
km s$^{-1}$ Mpc$^{-1}$.

An independent way to determine extragalactic distances comes from a
physical understanding  of the objects considered without the necessity
to use astronomical luminosity (or size) calibrators. The purely physical
calibration of the P-L relation of Cepheids has already been mentioned
(Sect.~2.1). For several X-ray clusters Sunyaev-Zel'dovich distances 
have become available. Three papers having appeared during the last
months give an (unweighted) mean of $H_0=58\pm 15$ (Holzapfel et al.~1997;
Lasenby \& Jones 1997; Myers et al.~1997). Rephaeli \& Yankovitch
(1997) have pointed out that the inclusion of relativistic effects reduces 
this value by $\sim 10$ units. Recent $H_0$ determinations from variable,
gravitationally lensed quasars give $H_0=59\pm 15$ (Keeton \& Kochanek
1997; Kundi\'c et al.~1997; Falco et al.~1997; Schechter et al.~1997).
The first interpretations of the CMB fluctuation spectrum suggest even
lower values (Lasenby \& Jones 1997; Lineweaver 1997).

In the light of this evidence values of $H_0$ below 45 and above 65 become
improbable. 

\acknowledgements

We have made use of the Lyon-Meudon Extragalactic Database (LEDA)
supplied by the LEDA team at the CRAL Observatoire de Lyon (France).
We thank S.M.G.~Hughes for his results prior to publication and
H.~B\"ohringer for the Virgo X-ray image from the ROSAT all-sky
survey. We are indebted to M.~Bershady for lending us his BCES
software. We thank the Referee for valuable comments. MF and GAT 
gratefully acknowledge the support of the Swiss
National Science Foundation. AS acknowledges support from NASA through
the Space Telescope Science Institute in the SNe$\,$Ia calibration
project using HST.

\clearpage
\begin{figure}
\epsscale{0.7}
\plotone{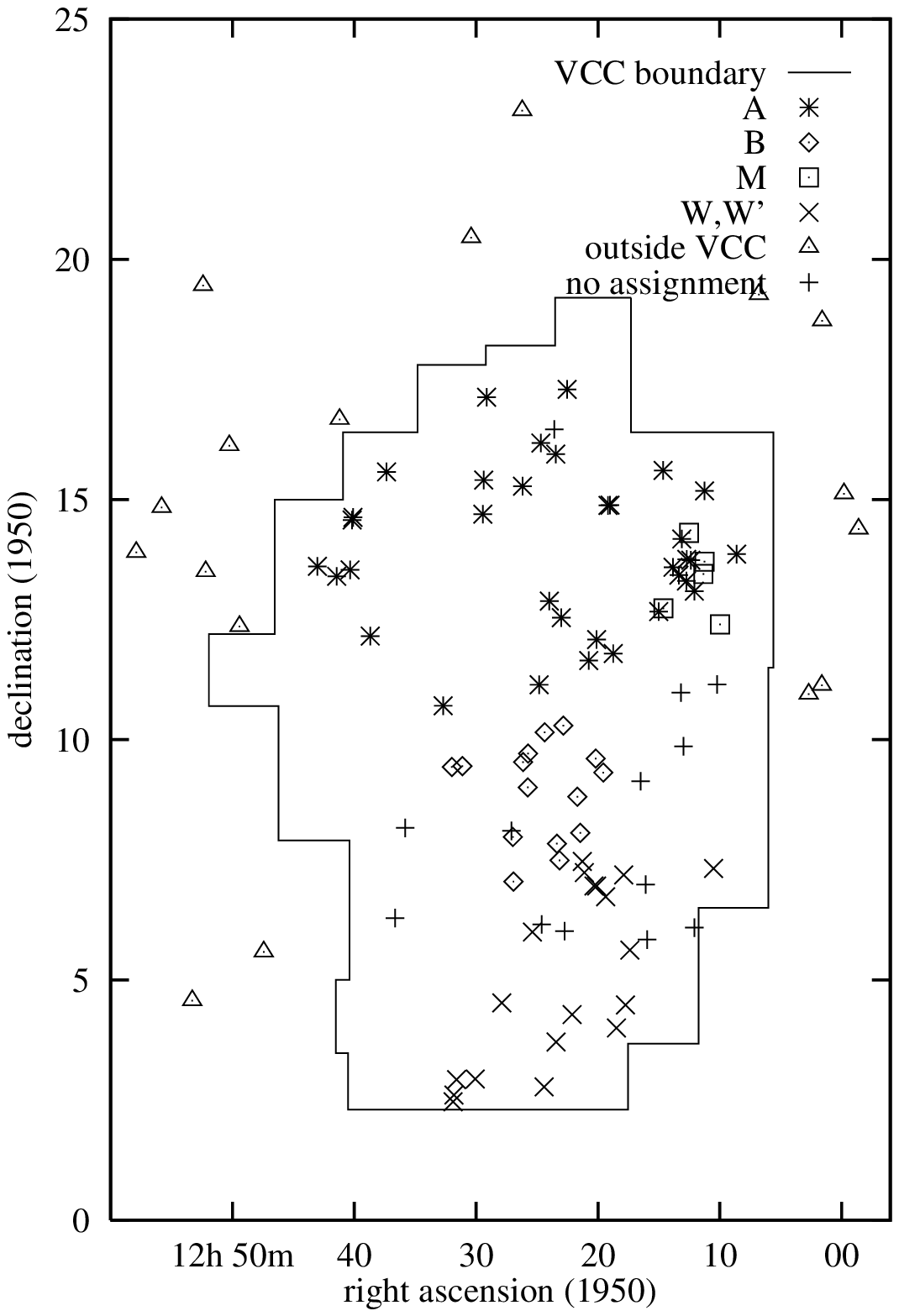}
\caption{Positions of the non-peculiar spirals of Table 3 with $i>45^{\circ}$
in the sky. The symbols denote the membership to a Virgo cluster subgroup
assigned by Binggeli et al.~(1993). Also shown is the boundary of the VCC
field (Binggeli et al.~1985).}
\end{figure}

\clearpage
\begin{figure}
\epsscale{0.7}
\plotone{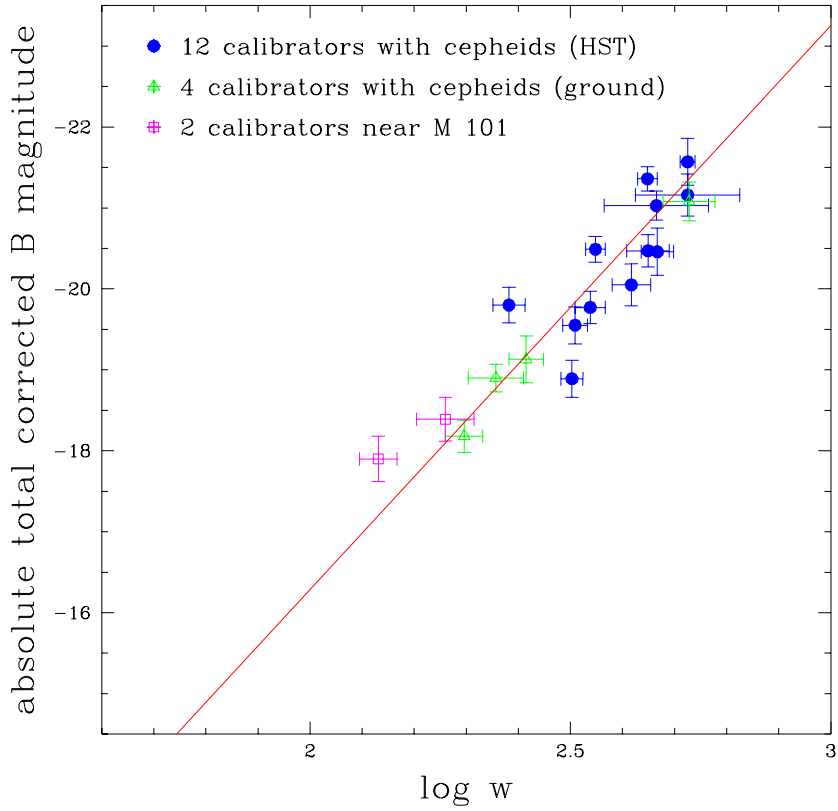}
\caption{Tully-Fisher relation for the calibrators of Table 2. Twelve
calibrators have Cepheid distances determined with HST, 4 have ground-based
Cepheid distances, 2 are members of the relatively tight M101 group without
individual Cepheid distances. The error bars show the total errors in
absolute magnitude und log linewidth which were used as weights for the 
regression. The line represents the adopted calibration of eq.~(3).}
\end{figure}

\clearpage
\begin{figure}
\epsscale{0.7}
\plotone{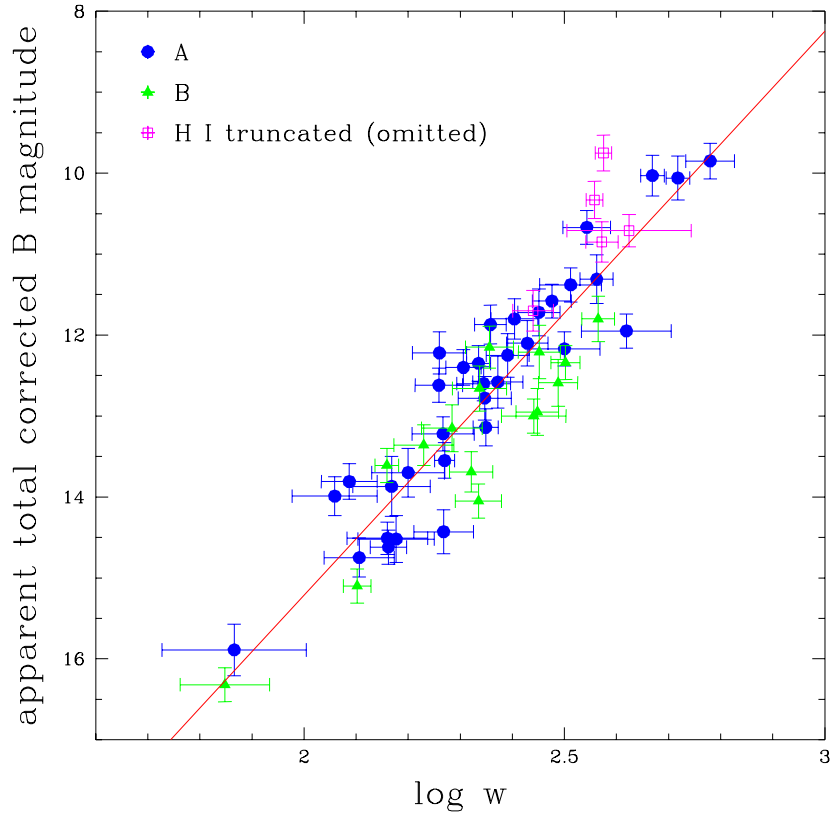}
\caption{Tully-Fisher relation for the \lq\lq fiducial Virgo 
sample\rq\rq . The error bars show the individual total errors in 
magnitude and log linewidth which served as weights for the calculation 
of the regression line [eq.(3)]. Four galaxies with H$\,$I-truncated disks are
also shown; they are systematically too bright for their line width and
were excluded from the Tully-Fisher analysis.}
\end{figure}

\clearpage
\begin{figure}
\epsscale{0.7}
\plotone{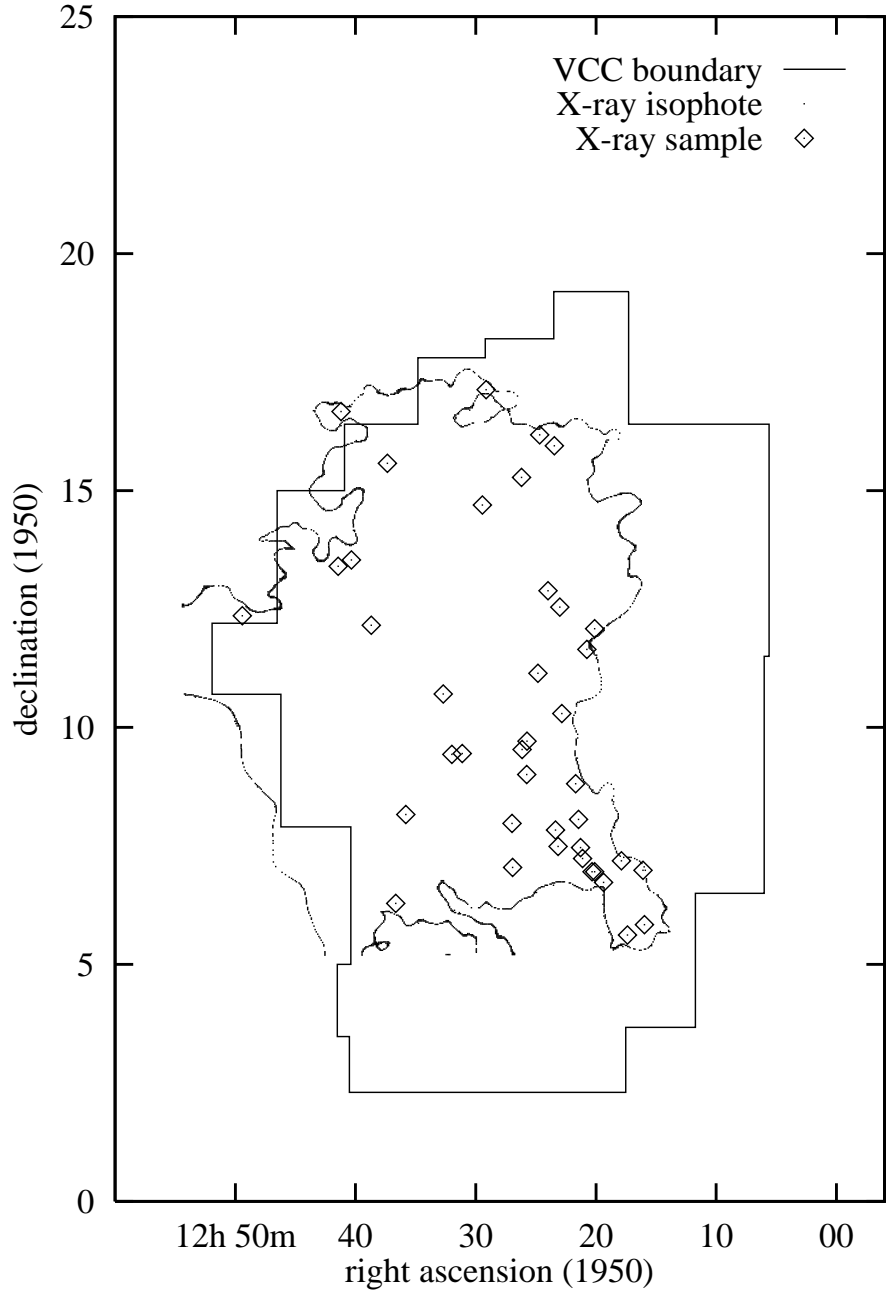}
\caption{Positions of the members of the X-ray sample (flag X in Table 3,
$i>45^{\circ}$) and its defining isophote from ROSAT data. Also shown is the
boundary of the VCC field (Binggeli et al.~1985).}
\end{figure}

\clearpage
\begin{figure}
\epsscale{0.7}
\plotone{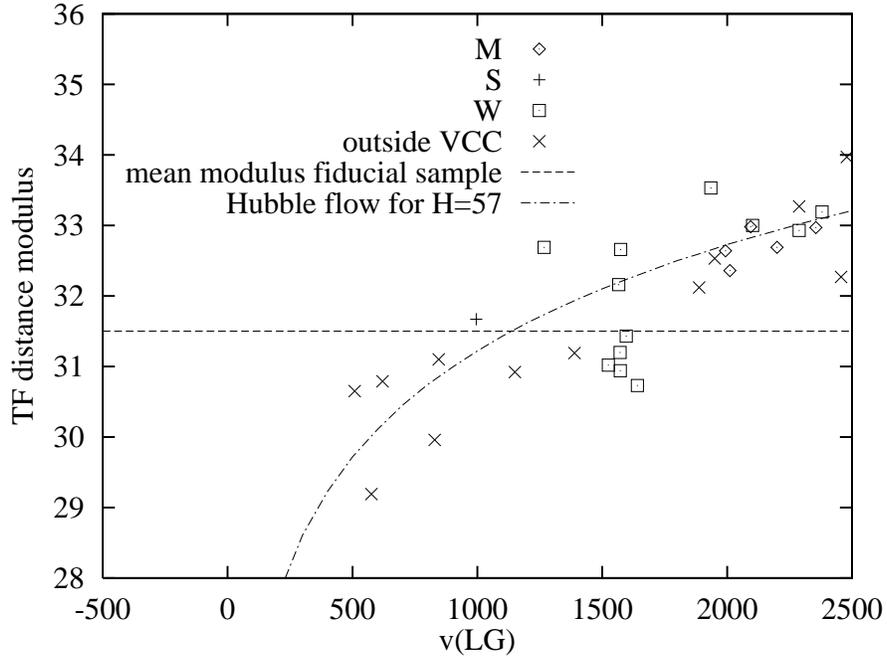}
\caption{Tully-Fisher distances of all galaxies in Table 3 outside the
fiducial sample and outside the X-ray sample as a function of redshift
[referred to the centroid of the Local Group using
eq.~(3)]. Memberships of different subgroups are denoted by different
symbols. There is a clear trend: the average distance is increasing
with increasing redshift.  This behaviour is compatible with a free
Hubble expansion (the dash-dotted line shows the cosmic expansion
assuming $H_0=57$).  The conclusion is that these galaxies do not
belong to the inner part of the Virgo cluster where internal motions
dominate the redshifts.}
\end{figure}

\clearpage
\begin{figure}
\epsscale{0.7}
\plotone{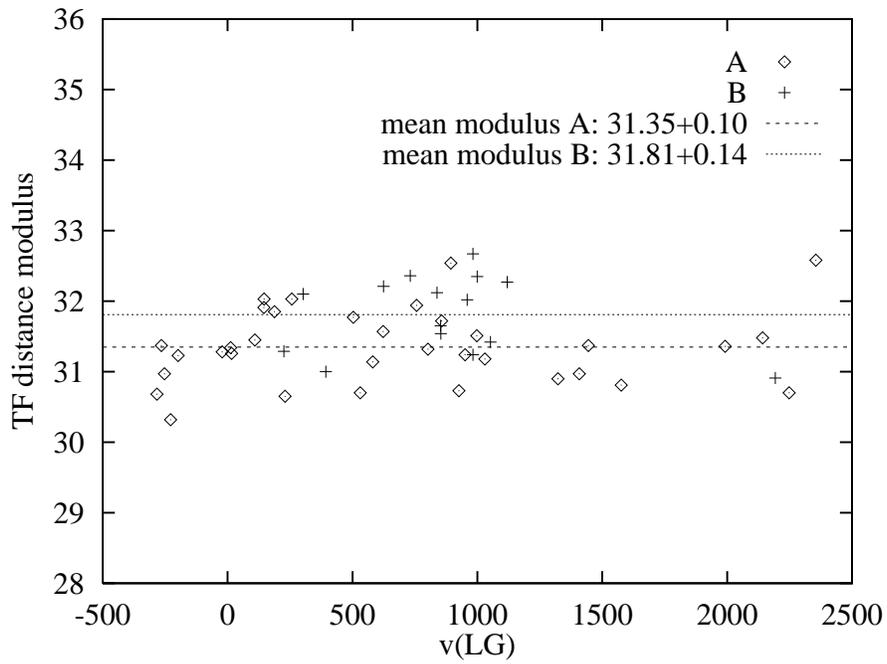}
\caption{Tully-Fisher distances for the members of the \lq\lq fiducial
sample\rq\rq ~as a function of redshift (referred to the centroid of the
Local Group) using eq.~(3). On average,  subgroup B lies about
0\fm 46 more distant than subgroup A. Obviously, the TF distances of the
cluster galaxies do not depend on redshift.}
\end{figure}

\clearpage
\begin{figure}
\epsscale{0.7}
\plotone{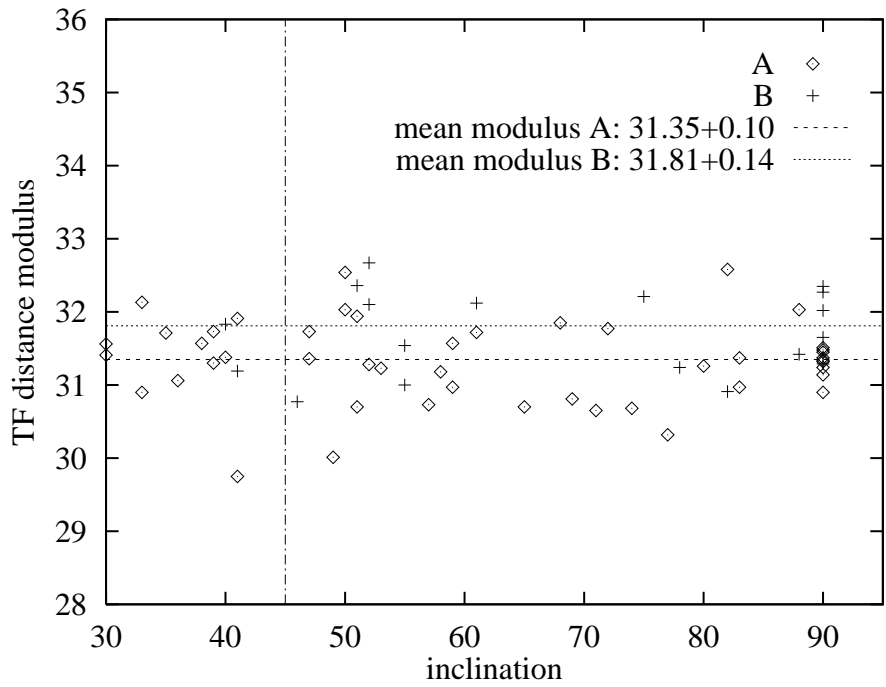}
\caption{TF distance of the fiducial sample as a function of inclination.
Also shown are the spirals with $30^{\circ} < i < 45^{\circ}$ (on the left
side of the vertical line).}
\end{figure}

\clearpage
\begin{figure}
\epsscale{0.7}
\plotone{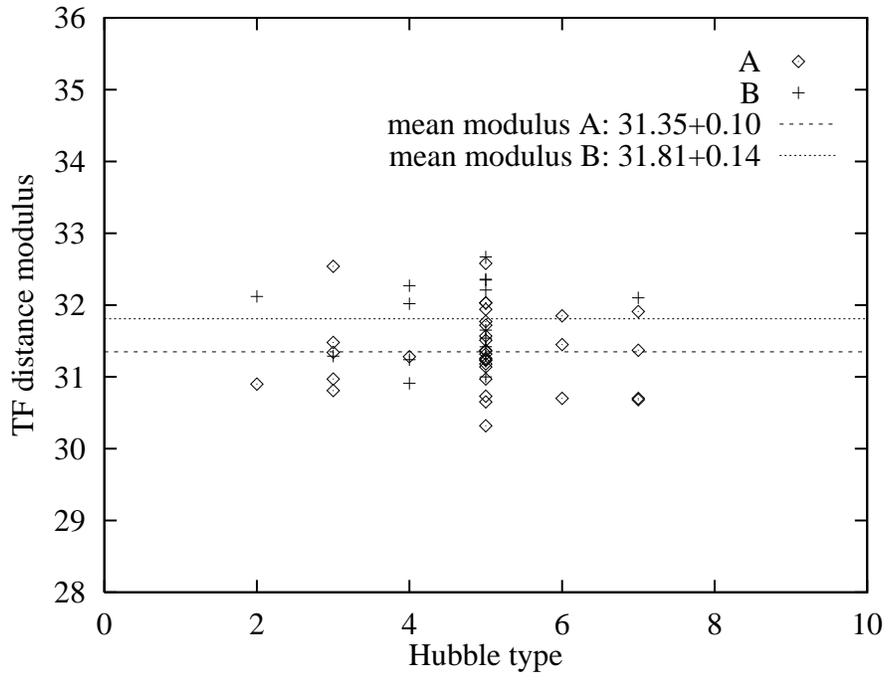}
\caption{TF distance of the fiducial sample as a function of Hubble type.}
\end{figure}

\clearpage
\begin{figure}
\epsscale{0.7}
\plotone{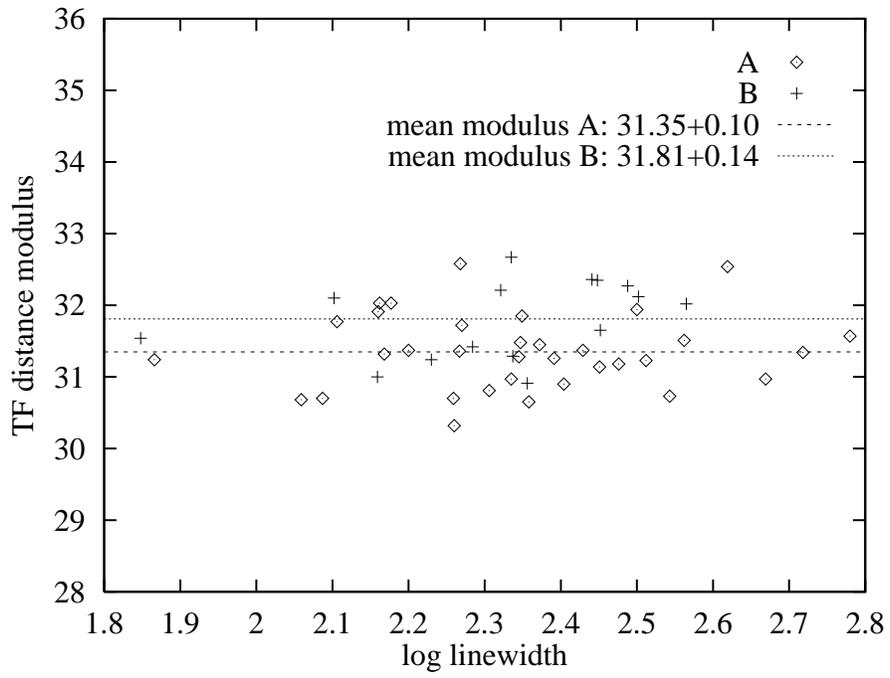}
\caption{TF distance of the fiducial sample as a function of line width.}
\end{figure}

\clearpage
\begin{figure}
\epsscale{0.7}
\plotone{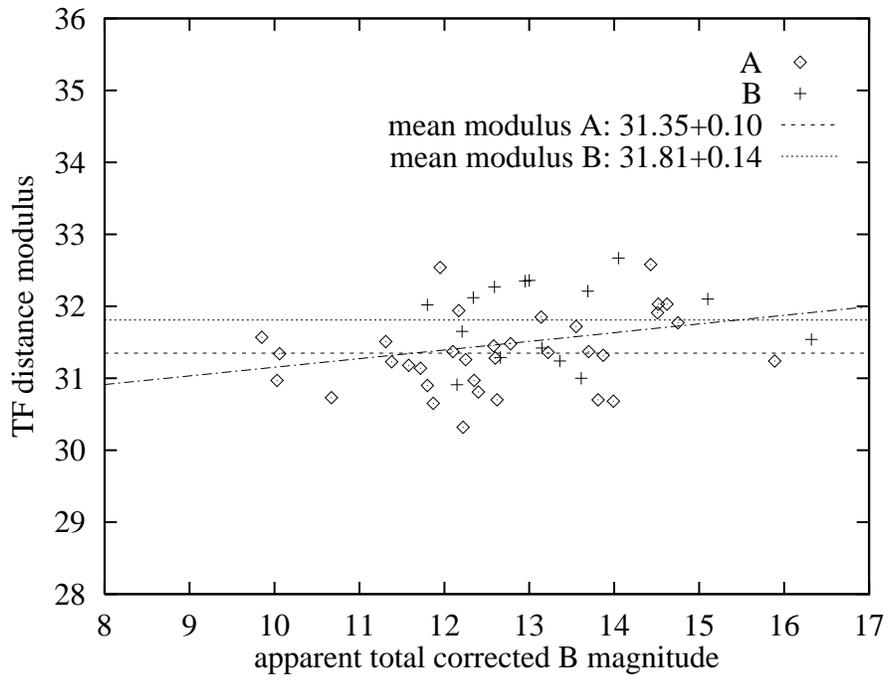}
\caption{TF distance of the fiducial sample as a function of apparent
magnitude. Due to the depth of the cluster of about $\pm 0\fm 4$, galaxies on
the near side are slightly brighter than those on the far side.}
\end{figure}

\clearpage
\begin{figure}
\epsscale{0.7}
\plotone{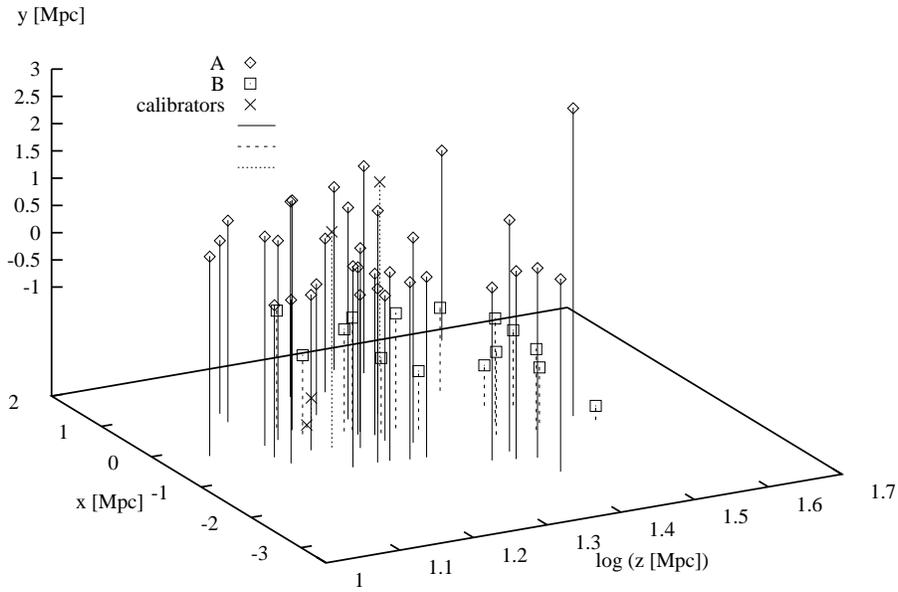}
\caption{3-D view of the Virgo cluster calculated from Tully-Fisher
distances. The $x$-$y$ plane corresponds to the sky as seen from the Sun (the
units being transformed into linear distances at the respective Virgo galaxy
distance), the z-axis shows the TF distance in logarithmic units $(\log
d_{\rm TF} {\rm [Mpc]})$. The Sun is to the left, M$\,$87 and the Sun is at
$x=y=0$. The calibrators among the Virgo galaxies are shown at their
respective Cepheid distance. The galaxy on the extreme right front
(at the largest $z$-distance) is IC 3033 (see Sect.~7).}
\end{figure}

\clearpage
\begin{figure}
\epsscale{0.7}
\plotone{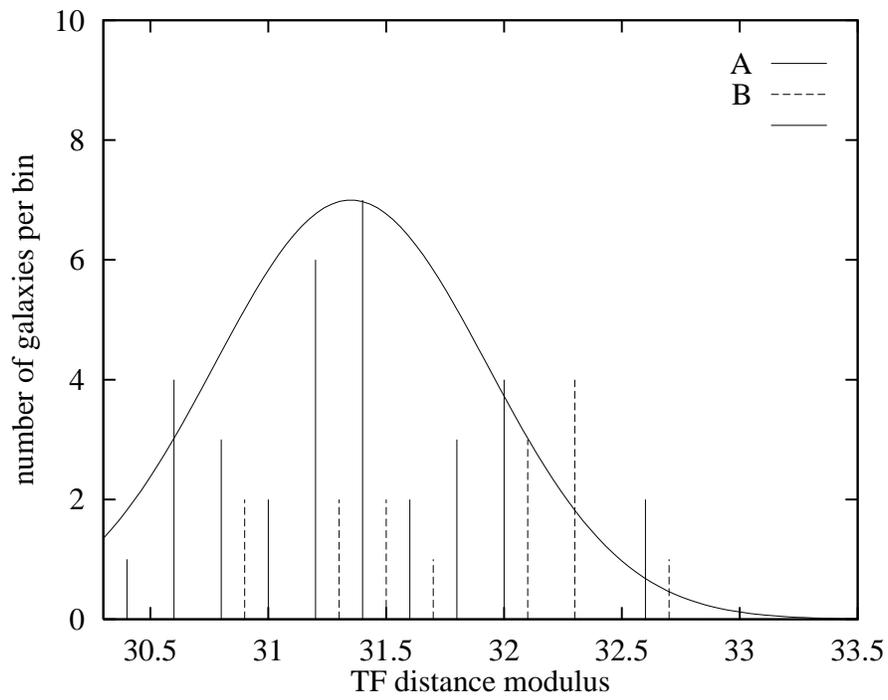}
\caption{Distribution of the individual distance moduli of subclusters 
A and B. As a crude approximation a Gaussians was fitted to the data of A; 
subcluster B has a bimodial rather than Gaussian distribution.}
\end{figure}

\clearpage
\begin{figure}
\epsscale{0.7}
\plotone{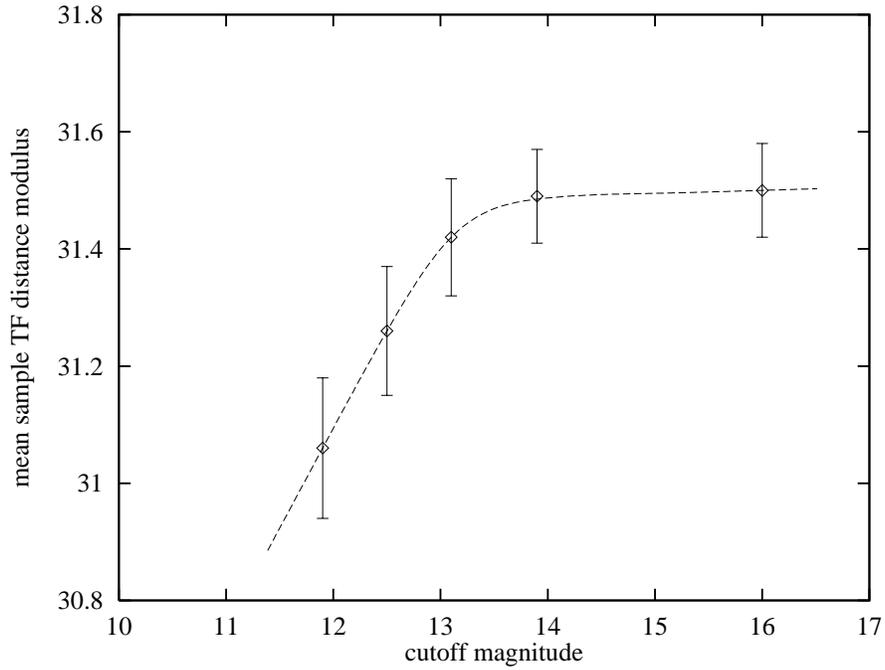}
\caption{The Teerikorpi cluster incompleteness bias for the fiducial
sample of the Virgo cluster. The apparent TF distance modulus is a
function of the apparent-magnitude cutoff of the sample. The
asymptotic value, which is the distance of the cluster, is only reached
if the sample of cluster galaxies is {\sl complete} and reaches deep
into the luminosity function (Sandage et al.~1995). The dashed line is a 
nonparametric fit through the data points using thin plate splines 
and regularisation.}
\end{figure}

\clearpage
\begin{figure}
\epsscale{0.7}
\plotone{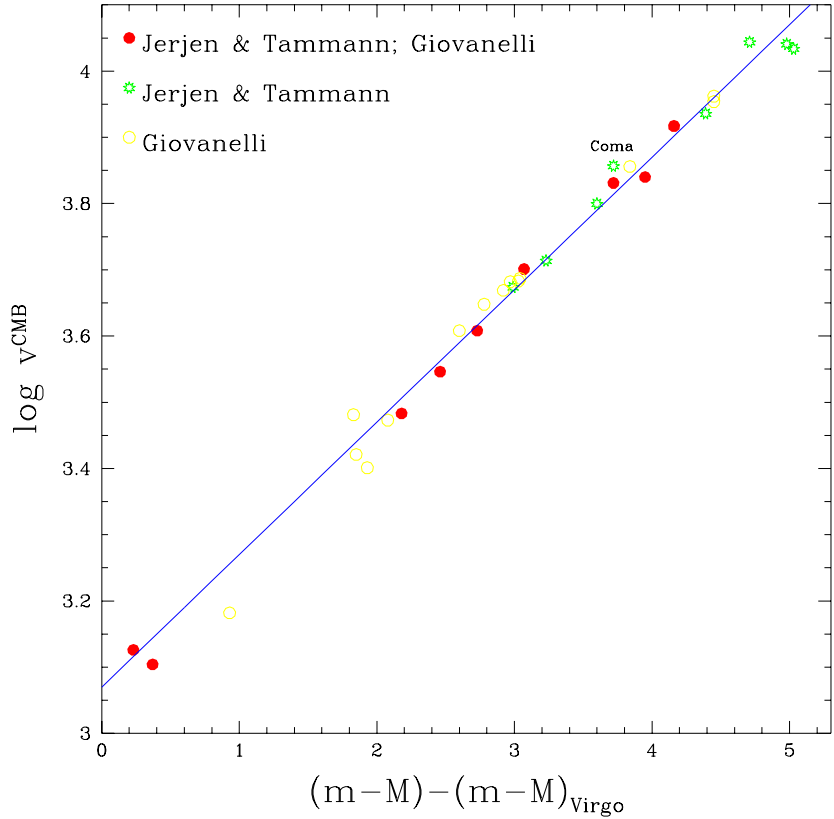}
\caption{Hubble diagram of 31 clusters with known relative
distances. The data were taken from Jerjen \& Tammann (1993;
asterisks) and Giovanelli (1996; open circles). Nine clusters are
listed in both sources (filled circles). The abscissa gives the
distance modulus relative to the Virgo cluster. The ordinate is the
log of the recession velocity referred to the CMB. For \lq\lq
local\rq\rq ~clusters with $v_0<3000\,$km s$^{-1}$ the velocities are
referred to the centroid of the Local Group and corrected for
Virgocentric infall, following the precepts of Federspiel et al.~(1994).}
\end{figure}

\end{document}


\singlespace
\begin{deluxetable}{lllcl}
\small
\tablewidth{0pc}
\tablecaption{Distance Determinations of LMC}
\tablehead{
\colhead{object} & \colhead{method} & \colhead{$(m-M)^0$} &
\colhead{zeropoint} & \colhead{Source}}

\startdata
ring of SN1987A & geometry & $18.56 \pm 0.05$ & -- & Panagia et al.~1996 \nl
RR Lyr & P-L-$\bigl [{{\rm Fe}\over {\rm H}}\bigr ]$ & $18.60\pm 0.11$ & Sandage 1993 & Walker 1993; Tammann 1996 \nl
red giants & tip of CMD & $18.43\pm 0.15$ & Sandage 1993 & Lee et al.~1993; Tammann 1996 \nl
Cepheids ($BVIJHK$) & Baade-Wesselink & $18.57\pm 0.15$ & -- & Laney \& Stobie 1992 \nl
Cepheids & P-L relation & $18.59\pm 0.15$ & Hyades & Sandage \& Tammann 1971 \nl
Cepheids & P-L relation & $18.47\pm 0.15$ & Pleiades & Feast \& Walker 1987 \nl
Cepheids & P-L relation & $18.57\pm 0.10$ & Pleiades & Feast 1995 \nl
Cepheids & P-L relation & $18.42\pm 0.11$ & Pleiades & B\"ohm-Vitense 1997 \nl
Cepheids & P-L relation & $18.70\pm 0.10$ & trig.~parallaxes & Feast \& Catchpole 1997 \nl
Mira variables & P-L relation & $18.54 \pm 0.10$ & trig.~parallaxes & van Leeuwen et al.~1997 \nl
Cepheids & $I$ Baade-Wesselink & $18.58\pm (0.02)$ & stellar radii & Di Benedetto 1997\nl
Cepheids & P-L relation & $18.57\pm 0.11$ & trig.~parallaxes & Madore \& Freedman 1997\nl
\hline
mean: & & $18.56\pm 0.03$ & & \nl
\enddata
\end{deluxetable}

\clearpage
\begin{deluxetable}{lcrrrccrrrrrr}
\small
\tablewidth{0pc}
\tablecaption{Data for the Calibrators}

\tablehead{
\colhead{Name} & \colhead{Hubble-} & \colhead{B$_T^{0,i}$} &
\colhead{A$^i$} & \colhead{A$^0$} & \colhead{$(m-M)$} &
\colhead{Source} & \colhead{$M_{\rm B}$} & \colhead{error} &
\colhead{$i_{\rm RC3}$} & \colhead{$\log w$} & \colhead{$\log w$} & \colhead{error}\\
\colhead{} & \colhead{type} & \colhead{mag} &
\colhead{mag} & \colhead{mag} & \colhead{mag} &
\colhead{} & \colhead{mag} & \colhead{mag} &
\colhead{$^{\circ}$} & \colhead{H\&R} & \colhead{LEDA} & \colhead{}\\
\colhead{(1)} & \colhead{(2)} & \colhead{(3)} &
\colhead{(4)} & \colhead{(5)} & \colhead{(6)} &
\colhead{(7)} & \colhead{(8)} & \colhead{(9)} &
\colhead{(10)} & \colhead{(11)} & \colhead{(12)} & \colhead{(13)}}

\startdata
N$\,$224 (M$\,$31) & 3  & 3.36 & 0.67 & 0.33 & 24.44 &1$\dagger$ & -21.08 &  0.24 & 78 & 2.739 & 2.728 & 0.050 \nl 
N$\,$300 & 5 & 8.49 & 0.21 & 0.02 & 26.67 &1 & -18.18 &  0.20 & 44 & 2.344 &  2.296 & 0.036 \nl
N$\,$598 (M$\,$33) & 5  & 5.73 & 0.35 & 0.19 & 24.63 &1 & -18.90 &  0.17 & 55 & 2.373 &  2.357 & 0.053 \nl
N$\,$925  & 5 & 10.04 & 0.38 & 0.27 & 29.84 & 2 & -19.80 & 0.22 & 58  & 2.395 & 2.382 & 0.031 \nl
N$\,$1365 & 4 & 9.94 & 0.38 & 0.00 & 31.30 & 3 & -21.36 & 0.15 & 59 & 2.655 & 2.648 & 0.019 \nl
N$\,$2090 & 5 & 11.52 & 0.47 & 0.00 & 30.41 & 4 & -18.89 & 0.30 & 63 & 2.526 & 2.503 & 0.021 \nl
N$\,$2403 & 5 & 8.38 & 0.38 & 0.17 & 27.51 & 1 & -19.13 &  0.29 & 62 & 2.484 &  2.415 & 0.033  \nl
N$\,$3031 (M$\,$81) & 3 & 7.34  & 0.39 & 0.16 & 27.80 & 5 & -20.46 & 0.29 & 65 & 2.697 &  2.667 & 0.031 \nl
N$\,$3351 (M$\,$95) & 3 & 10.24 & 0.24 & 0.05 & 30.01 & 6 & -19.77 & 0.20 & 50 & 2.546 & 2.538 & 0.029 \nl
N$\,$3368 (M$\,$96) & 2 & 9.85 & 0.20 & 0.06 & 30.32 & 7 & -20.47 & 0.20 & 50 & 2.656 & 2.649 & 0.041 \nl
N$\,$3621 & 5 & 9.40 & 0.36 & 0.42 & 28.95 & 8 & -19.55 & 0.23 & 57 & 2.518 & 2.509 & 0.024 \nl
N$\,$4321 (M$\,$100) & 5 & 9.88 & 0.11 & 0.06 & 31.04 & 9 & -21.16 & 0.26 & \phantom{2}29$^{\ast}$ & 2.733 & 2.725 & 0.100 \nl
N$\,$4536 & 4 & 10.62 & 0.54 & 0.00 & 31.11 & 10  & -20.49 & 0.16 & 66 & 2.546 & 2.548 & 0.019 \nl
N$\,$4639 & 3 & 11.95 & 0.23 & 0.06 & 32.00 & 11 & -20.05 & 0.26 & 50 & 2.626 & 2.617 & 0.037 \nl
N$\,$5204 & 7 & 11.43 & 0.30 & 0.00 & 29.34 & 12 & -17.87 & 0.28 & 57 & 2.146 & 2.131 & 0.036 \nl
N$\,$5457 (M$\,$101) & 5 & 8.31 & 0.00 & 0.00 & 29.34 &12 & -21.03 & 0.18 & \phantom{0}22$^{\ast}$ & 2.680 & 2.665 & 0.100 \nl
N$\,$5585 & 7 & 10.94 & 0.26 & 0.00 & 29.34 & 12 & -18.36 & 0.27 & 52 & 2.290 & 2.260 & 0.055 \nl
N$\,$7331 & 3 & 9.39 & 0.61 & 0.35 & 30.94 & 13 & -21.57 & 0.25 & 75 & 2.740 & 2.725 & 0.014 \nl
\enddata
\end{deluxetable}

\clearpage
\begin{deluxetable}{llcccrrrclll}
\small
\tablewidth{0pc}
\tablecaption{Data for Virgo Cluster Galaxies}
\tablehead{
\colhead{Name} & \colhead{sub-} & \colhead{Hubble-} & \colhead{$B_T^{0,i}$} & 
\colhead{error} & \colhead{$A^i_{\rm RC3}$} & \colhead{$A^i_{\rm RSA}$} &
\colhead{$A^0_{\rm RC3}$} & \colhead{$i_{\rm RC3}$} & \colhead{$\log w$} &
\colhead{$\log w$} & \colhead{error} \\
\colhead{} & \colhead{group} & \colhead{type} & \colhead{mag} & \colhead{mag} &
\colhead{mag} & \colhead{mag}  & \colhead{} &
\colhead{$^{\circ}$} & \colhead{H\&R} & \colhead{LEDA} & \colhead{} \\
\colhead{(1)} & \colhead{(2)} & \colhead{(3)} &
\colhead{(4)} & \colhead{(5)} & \colhead{(6)} &
\colhead{(7)} & \colhead{(8)} & \colhead{(9)} &
\colhead{(10)} & \colhead{(11)} & \colhead{(12)} }
\startdata
\multicolumn{11}{l}{Members of \lq\lq big A\rq\rq ~or \lq\lq B 
($r=2.4^{\circ}$)\rq\rq , $i>45^{\circ}$, (\lq\lq fiducial sample\rq\rq)
\hfill} \nl
I3033  & A  &  5 &  14.62 &   .21 &   .27 &   .44 &   .03 &  50 &   2.276\phantom{$^{\ast\ast}$} &  2.162 & 0.035 \nl 
N4192  & A  &  3 &  10.03 &   .27 &   .76 &  1.17 &   .16 &  83 &   2.675\phantom{$^{\ast\ast}$} &  2.669 & 0.023 \nl 
U7249  & A  &  5 &  14.75 &   .24 &   .66 &   .67 &   .02 &  72 &   2.130\phantom{$^{\ast\ast}$} &  2.106 & 0.067 \nl 
I3059  & A  &  7 &  14.51 &   .20 &   .17 &   .39 &   .02 &  41 &   2.187$^{\ast\ast}$ &  2.160 & 0.078 \nl 
I3066  & A  &  5 &  14.52 &   .29 &  1.03 &   .87 &   .02 &  88 &   2.194$^{\ast\ast}$ &  2.177 & 0.073 \nl 
N4206  & A  &  5 &  11.72 &   .30 &  1.08 &   .87 &   .02 &  90 &   2.450\phantom{$^{\ast\ast}$} &  2.451 & 0.041 \nl 
N4212  & A  &  5 &  11.38 &   .22 &   .32 &   .46 &   .14 &  53 &   2.558\phantom{$^{\ast\ast}$} &  2.512 & 0.059 \nl 
N4216  & A  &  3 &  10.06 &   .27 &   .91 &  1.31 &   .02 &  90 &   2.732\phantom{$^{\ast\ast}$} &  2.718 & 0.023 \nl 
N4222  & A  &  6 &  12.58 &   .33 &  1.26 &   .87 &   .02 &  90 &   2.367\phantom{$^{\ast\ast}$} &  2.372 & 0.048 \nl 
N4237  & A  &  5 &  12.17 &   .21 &   .29 &   .45 &   .05 &  51 &   2.540\phantom{$^{\ast\ast}$} &  2.500 & 0.068 \nl 
I3105  & A  &  7 &  13.99 &   .25 &   .70 &   .73 &   .04 &  74 &   2.068\phantom{$^{\ast\ast}$} &  2.059 & 0.082 \nl 
N4294  & A  &  5 &  11.87 &   .24 &   .63 &   .65 &   .03 &  70 &   2.362\phantom{$^{\ast\ast}$} &  2.358 & 0.031 \nl 
N4298  & A  &  5 &  11.58 &   .22 &   .38 &   .50 &   .08 &  57 &   2.500\phantom{$^{\ast\ast}$} &  2.476 & 0.037 \nl 
N4302  & A  &  5 &  11.31 &   .31 &  1.11 &   .87 &   .08 &  90 &   2.567\phantom{$^{\ast\ast}$} &  2.562 & 0.032 \nl 
N4307  & B  &  4 &  11.80 &   .28 &   .98 &   .87 &   .00 &  90 &   2.588\phantom{$^{\ast\ast}$} &  2.565 & 0.032 \nl 
N4313  & A, X &  2 &  11.80 &   .25 &   .73 &  1.22 &   .01 &  90 &   2.422\phantom{$^{\ast\ast}$} &  2.404 & 0.045 \nl 
N4316  & B  &  4 &  12.59 &   .29 &  1.03 &   .87 &   .00 &  90 &   2.500\phantom{$^{\ast\ast}$} &  2.488 & 0.037 \nl 
N4330  & A, X &  7 &  12.10 &   .28 &   .97 &   .87 &   .02 &  82 &   2.464\phantom{$^{\ast\ast}$} &  2.429 & 0.040 \nl 
N4353  & B, X &  5 &  14.05 &   .21 &   .30 &   .46 &   .00 &  52 &   2.352\phantom{$^{\ast\ast}$} &  2.335 & 0.045 \nl 
N4356  & B, X &  5 &  12.95 &   .29 &  1.06 &   .87 &   .00 &  90 &   2.486\phantom{$^{\ast\ast}$} &  2.448 & 0.041 \nl 
I3298  & A  &  5 &  14.43 &   .27 &   .92 &   .82 &   .01 &  81 &   2.299\phantom{$^{\ast\ast}$} &  2.268 & 0.057 \nl 
N4380  & B, X &  2 &  12.34 &   .21 &   .32 &   .78 &   .00 &  60 &   2.524\phantom{$^{\ast\ast}$} &  2.502 & 0.028 \nl 
I3311  & A, X &  5 &  13.70 &   .31 &  1.12 &   .87 &   .06 &  90 &   2.260\phantom{$^{\ast\ast}$} &  2.200 & 0.070 \nl 
I3322A & B, X &  5 &  12.21 &   .33 &  1.33 &   .87 &   .00 &  90 &   2.442\phantom{$^{\ast\ast}$} &  2.452 & 0.040 \nl 
I3322  & B, X &  5 &  13.15 &   .29 &  1.03 &   .87 &   .00 &  88 &   2.338\phantom{$^{\ast\ast}$} &  2.284 & 0.058 \nl 
N4396  & A, X &  5 &  12.22 &   .26 &   .80 &   .75 &   .04 &  77 &   2.279\phantom{$^{\ast\ast}$} &  2.260 & 0.052 \nl 
N4413  & A, X &  4 &  12.60 &   .21 &   .28 &   .45 &   .11 &  51 &   2.316\phantom{$^{\ast\ast}$} &  2.345 & 0.052 \nl 
V 952  & B, X &  5 &  16.32 &   .21 &   .33 &   .47 &   .00 &  54 &   1.891$^{\ast\ast}$ &  1.848 & 0.086 \nl 
I3365  & A, X &  6 &  13.81 &   .23 &   .50 &   .58 &   .03 &  64 &   2.155\phantom{$^{\ast\ast}$} &  2.087 & 0.054 \nl 
I3371  & A, X &  5 &  13.88 &   .37 &  1.54 &   .87 &   .02 &  90 &   2.186$^{\ast}$\phantom{$^{\ast}$} &  2.168 & 0.075 \nl 
N4445  & B, X &  3 &  12.66 &   .28 &  1.01 &  1.33 &   .00 &  90 &   2.356\phantom{$^{\ast\ast}$} &  2.337 & 0.052 \nl 
U7590  & B, X &  4 &  13.36 &   .25 &   .76 &   .74 &   .00 &  78 &   2.241\phantom{$^{\ast\ast}$} &  2.230 & 0.058 \nl 
N4451  & B, X &  5 &  13.00 &   .21 &   .29 &   .45 &   .00 &  51 &   2.498\phantom{$^{\ast\ast}$} &  2.441 & 0.062 \nl 
I3392  & A, X &  3 &  12.40 &   .23 &   .51 &   .93 &   .08 &  69 &   2.340\phantom{$^{\ast\ast}$} &  2.306 & 0.034 \nl 
I3414  & B, X &  5 &  13.61 &   .21 &   .33 &   .47 &   .00 &  54 &   2.182\phantom{$^{\ast\ast}$} &  2.159 & 0.023 \nl 
N4466  & B, X &  5 &  13.69 &   .25 &   .73 &   .71 &   .00 &  75 &   2.373\phantom{$^{\ast\ast}$} &  2.321 & 0.041 \nl 
N4498  & A, X &  5 &  12.35 &   .22 &   .41 &   .52 &   .03 &  59 &   2.373\phantom{$^{\ast\ast}$} &  2.335 & 0.023 \nl 
I 797  & A, X &  5 &  13.22 &   .21 &   .24 &   .42 &   .09 &  47 &   2.292\phantom{$^{\ast\ast}$} &  2.267 & 0.060 \nl 
N4501  & A, X &  5 &   9.85 &   .22 &   .41 &   .52 &   .10 &  59 &   2.793\phantom{$^{\ast\ast}$} &  2.780 & 0.047 \nl 
N4522  & B, X &  4 &  12.15 &   .26 &   .84 &   .78 &   .00 &  81 &   2.377\phantom{$^{\ast\ast}$} &  2.356 & 0.046 \nl 
I3517  & B, X &  7 &  15.10 &   .21 &   .28 &   .46 &   .00 &  51 &   2.130\phantom{$^{\ast\ast}$} &  2.102 & 0.027 \nl 
V1605  & A, X &  5 &  15.89 &   .32 &  1.26 &   .87 &   .00 &  90 &   1.900$^{\ast\ast}$ &  1.866 & 0.139 \nl 
N4595  & A, X &  7 &  12.62 &   .21 &   .26 &   .45 &   .03 &  50 &   2.286\phantom{$^{\ast\ast}$} &  2.259 & 0.046 \nl 
N4607  & A, X &  3 &  12.78 &   .28 &   .88 &  1.29 &   .09 &  90 &   2.365\phantom{$^{\ast\ast}$} &  2.347 & 0.051 \nl 
N4633  & A  &  6 &  13.14 &   .23 &   .57 &   .62 &   .04 &  68 &   2.338\phantom{$^{\ast\ast}$} &  2.349 & 0.024 \nl 
N4634  & A  &  5 &  12.25 &   .27 &   .87 &   .79 &   .04 &  80 &   2.444\phantom{$^{\ast\ast}$} &  2.391 & 0.041 \nl 
N4639  & A, X &  3 &  11.95 &   .21 &   .23 &   .66 &   .06 &  49 &   2.626\phantom{$^{\ast\ast}$} &  2.619 & 0.086 \nl 
N4654  & A, X &  5 &  10.67 &   .22 &   .36 &   .49 &   .07 &  56 &   2.571\phantom{$^{\ast\ast}$} &  2.543 & 0.046 \nl 
I3742  & A  &  5 &  13.55 &   .22 &   .43 &  .54 &   .07 &  61 &   2.316\phantom{$^{\ast\ast}$} &  2.270 & 0.019 \nl 
\hline
\multicolumn{11}{l}{H$\,$I-truncated galaxies (Guhathakurta et al.~1988)}\nl
N4388 & A, X &  2 &  10.85 &   .27 &   .79 & 1.29 &   .12 &  90 &   2.566\phantom{$^{\ast\ast}$} &  2.572 & 0.031 \nl 
N4402 & A, X &  3 &  11.70 &   .27 &   .76 & 1.17 &   .13 &  83 &   2.446\phantom{$^{\ast\ast}$} &  2.440 & 0.039 \nl 
N4438 & A, X &  3 &  10.33 &   .24 &   .59 & 1.01 &   .10 &  73 &   2.606\phantom{$^{\ast\ast}$} &  2.558 & 0.016 \nl 
N4450 & A  &  2 &  10.71 &   .20 &   .16 &  .60 &   .03 &  44 &   2.534\phantom{$^{\ast\ast}$} &  2.624 & 0.119 \nl 
N4569 & A, X &  2 &   9.75 &   .22 &   .42 &  .89 &   .09 &  68 &   2.591\phantom{$^{\ast\ast}$} &  2.575 & 0.016 \nl 
\hline
\multicolumn{11}{l}{peculiar or interacting galaxies (Binggeli et al.~1985)}\nl
I3044  & A  &  5 &  13.58 &   .24 &   .57 &  .61 &   .09 &  68 &   2.190\phantom{$^{\ast\ast}$} &  2.154 & 0.052 \nl 
N4568$^1$& A, X &  5 &  11.13 &   .23 &   .54 &  .60 &   .01 &  66 &   2.589\phantom{$^{\ast\ast}$} &  2.565 & 0.016 \nl
\hline
\multicolumn{11}{l}{members of A, B with $30^{\circ}<i<45^{\circ}$}\nl
U7200  & A  &  8 &  14.13 &   .20 &   .12 &  .18 &   .05 &  40 &   0.000$^{\dagger}$\phantom{$^{\ast}$} &  0.000 & 0.000 \nl
V 132  & A  &  7 &  16.37 &   .20 &   .17 &  .39 &   .02 &  41 &   1.677$^{\ast}$\phantom{$^{\ast}$} &  1.585 & 0.147 \nl
N4254  & A  &  5 &  10.24 &   .20 &   .09 &  .33 &   .11 &  30 &   2.736\phantom{$^{\ast\ast}$} &  2.702 & 0.260 \nl
N4321  & A, X &  4 &   9.89 &   .20 &   .10 &  .34 &   .06 &  32 &   2.747\phantom{$^{\ast\ast}$} &  2.679 & 0.232 \nl
I3258  & A, X &  5 &  13.40 &   .20 &   .09 &  .33 &   .08 &  30 &   2.310\phantom{$^{\ast\ast}$} &  2.270 & 0.192 \nl
N4351  & A, X &  5 &  12.76 &   .21 &   .26 &  .43 &   .01 &  44 &   2.236\phantom{$^{\ast\ast}$} &  2.138 & 0.041 \nl
N4390  & A, X &  4 &  13.18 &   .20 &   .10 &  .34 &   .00 &  32 &   2.352\phantom{$^{\ast\ast}$} &  2.384 & 0.189 \nl
N4411A & B, X &  5 &  13.38 &   .20 &   .03 &  .30 &   .00 &  17$^2$ &   2.405\phantom{$^{\ast\ast}$} &  2.484 & 0.410 \nl
I3476  & A, X &  5 &  12.95 &   .21 &   .12 &  .35 &   .12 &  34 &   2.405\phantom{$^{\ast\ast}$} &  2.355 & 0.168 \nl
I3483$^3$  & A, X &  3 &  15.52 &   .20 &   .17 &  .59 &   .01 &  42 &   2.348\phantom{$^{\ast\ast}$} &  2.276 & 0.097 \nl 
N4519  & B, X &  5 &  12.18 &   .20 &   .17 &  .38 &   .00 &  40 &   2.446\phantom{$^{\ast\ast}$} &  2.484 & 0.142 \nl 
N4535  & B, X &  5 &  10.37 &   .20 &   .22 &  .41 &   .00 &  44 &   2.606\phantom{$^{\ast\ast}$} &  2.591 & 0.107 \nl
I3520  & A, X &  6 &  14.68 &   .21 &   .24 &  .42 &   .08 &  42 &   2.190\phantom{$^{\ast\ast}$} &  2.109 & 0.044 \nl
N4540  & A, X &  6 &  12.23 &   .20 &   .16 &  .38 &   .05 &  39 &   2.318\phantom{$^{\ast\ast}$} &  2.412 & 0.134 \nl 
N4548  & A, X &  3 &  10.75 &   .20 &   .14 &  .56 &   .07 &  38 &   2.663\phantom{$^{\ast\ast}$} &  2.612 & 0.164 \nl
N4571  & A, X &  5 &  11.69 &   .20 &   .08 &  .32 &   .06 &  27$^4$ &   2.548\phantom{$^{\ast\ast}$} &  2.534 & 0.271 \nl 
N4579  & A, X &  2 &  10.21 &   .21 &   .12 &  .56 &   .15 &  39 &   2.787\phantom{$^{\ast\ast}$} &  2.753 & 0.170 \nl 
U7795  & B, X &  8 &  14.54 &   .20 &   .14 &  .19 &   .00 &  41 &   2.064\phantom{$^{\ast\ast}$} &  2.053 & 0.059 \nl 
N4647  & A, X &  5 &  11.75 &   .20 &   .15 &  .37 &   .04 &  38 &   2.505\phantom{$^{\ast\ast}$} &  2.507 & 0.157 \nl
N4689  & A, X &  5 &  11.41 &   .20 &   .14 &  .36 &   .05 &  36 &   2.580\phantom{$^{\ast\ast}$} &  2.482 & 0.170 \nl
\hline
\multicolumn{11}{l}{galaxies outside A or B but within X-ray contours:} \nl
N4252  & Z, X &  5 &  13.96 &   .27 &   .92 &  .82 &   .00 &  81 &   2.212$^{\ast\ast}$ &  2.196 & 0.067 \nl 
V 297  & Z, X &  5 &  14.00 &   .31 &  1.22 &  .87 &   .00 &  90 &   2.239$^{\ast}$\phantom{$^{\ast}$} &  2.224 & 0.066 \nl
N4273  & W, X &  5 &  12.09 &   .21 &   .30 &  .46 &   .00 &  52 &   2.558$^{\ast\ast}$ &  2.558 & 0.066 \nl 
V 415  & W, X &  7 &  14.57 &   .21 &   .37 &  .52 &   .00 &  58 &   2.163$^{\ast\ast}$ &  2.111 & 0.036 \nl
U7423  & X, X &  7 &  15.23 &   .21 &   .32 &  .48 &   .00 &  54 &   2.212\phantom{$^{\ast\ast}$} &  2.231 & 0.022 \nl 
I3225  & W, X &  6 &  13.79 &   .24 &   .71 &  .70 &   .00 &  73 &   2.353$^{\ast\ast}$ &  2.333 & 0.038 \nl 
I3229  & X, X &  5 &  14.39 &   .26 &   .84 &  .77 &   .00 &  78 &   2.161\phantom{$^{\ast\ast}$} &  2.120 & 0.077 \nl 
N4343  & X, X &  3 &  12.37 &   .25 &   .75 & 1.15 &   .00 &  82 &   2.534\phantom{$^{\ast\ast}$} &  2.535 & 0.031 \nl 
I3259  & X, X &  5 &  13.85 &   .21 &   .39 &  .51 &   .00 &  58 &   2.305\phantom{$^{\ast\ast}$} &  2.288 & 0.015 \nl 
N4470$^3$  & Z, X &  5 &  12.81 &   .20 &   .21 &  .40 &   .00 &  44 &   2.324\phantom{$^{\ast\ast}$} &  2.291 & 0.080 \nl
I 800$^3$  & Z, X &  5 &  13.81 &   .21 &   .21 &  .40 &   .05 &  44 &   2.258\phantom{$^{\ast\ast}$} &  2.137 & 0.058 \nl
U7802  & Z, X &  5 &  13.86 &   .30 &  1.11 &  .87 &   .00 &  90 &   2.238\phantom{$^{\ast\ast}$} &  2.216 & 0.068 \nl 
N4591  & Z, X &  3 &  13.60 &   .22 &   .41 &  .83 &   .00 &  63 &   2.603\phantom{$^{\ast\ast}$} &  2.538 & 0.021 \nl 
N4651  & Y, X &  5 &  11.09 &   .21 &   .27 &  .44 &   .03 &  50 &   2.718\phantom{$^{\ast\ast}$} &  2.685 & 0.088 \nl 
N4746  & Y, X &  3 &  12.57 &   .25 &   .72 & 1.13 &   .07 &  80 &   2.543\phantom{$^{\ast\ast}$} &  2.535 & 0.029 \nl 
\multicolumn{11}{l}{galaxies outside A or B and outside X-ray contours}\nl
I 755  & Y  &  3 &  12.79 &   .31 &  1.21 &  1.33 &   .00 &  90 &   2.316\phantom{$^{\ast\ast}$} &  2.304 & 0.056 \nl 
U7016  & Y  &  2 &  13.60 &   .26 &   .81 &  1.31 &   .01 &  90 &   0.000$^{\dagger}$\phantom{$^{\ast}$} &  0.000 & 0.000 \nl 
N4064  & Y  &  5 &  11.60 &   .24 &   .60 &   .63 &   .02 &  69 &   2.318\phantom{$^{\ast\ast}$} &  2.299 & 0.032 \nl 
N4067  & Y  &  3 &  13.12 &   .20 &   .19 &   .62 &   .00 &  45 &   2.587\phantom{$^{\ast\ast}$} &  2.556 & 0.107 \nl 
N4082  & Y  &  5 &  14.69 &   .20 &   .00 &   .28 &   .19 &   0 &   0.000$^{\dagger}$\phantom{$^{\ast}$} &  0.000 & 0.000 \nl 
U7133  & Y  &  7 &  15.00 &   .21 &   .26 &   .45 &   .03 &  50 &   2.354\phantom{$^{\ast\ast}$} &  2.387 & 0.059 \nl 
U7209  & M  &  3 &  13.12 &   .21 &   .23 &   .66 &   .03 &  49 &   2.549\phantom{$^{\ast\ast}$} &  2.515 & 0.078 \nl 
N4178  & Z  &  5 &  11.22 &   .24 &   .68 &   .68 &   .00 &  72 &   2.491\phantom{$^{\ast\ast}$} &  2.454 & 0.024 \nl 
N4180  & W  &  3 &  12.89 &   .23 &   .58 &   .99 &   .00 &  73 &   2.645\phantom{$^{\ast\ast}$} &  2.625 & 0.014 \nl 
N4189  & M  &  5 &  12.27 &   .21 &   .21 &   .40 &   .03 &  44 &   2.672\phantom{$^{\ast\ast}$} &  2.586 & 0.113 \nl 
N4193  & M  &  5 &  12.68 &   .22 &   .48 &   .56 &   .02 &  63 &   2.602\phantom{$^{\ast\ast}$} &  2.576 & 0.025 \nl 
N4197  & Z  &  6 &  12.29 &   .30 &  1.15 &   .87 &   .00 &  90 &   2.442\phantom{$^{\ast\ast}$} &  2.427 & 0.043 \nl 
I3061  & M  &  5 &  13.13 &   .31 &  1.06 &   .87 &   .14 &  90 &   2.509\phantom{$^{\ast\ast}$} &  2.470 & 0.038 \nl 
N4207  & Z  &  6 &  13.05 &   .22 &   .43 &   .54 &   .00 &  60 &   2.393\phantom{$^{\ast\ast}$} &  2.346 & 0.022 \nl 
I3074  & Z  &  7 &  13.58 &   .30 &  1.15 &   .87 &   .00 &  90 &   2.371\phantom{$^{\ast\ast}$} &  2.335 & 0.052 \nl 
I3099  & M  &  6 &  13.68 &   .32 &  1.21 &   .87 &   .04 &  90 &   2.398\phantom{$^{\ast\ast}$} &  2.343 & 0.051 \nl 
I 776  & Z  &  6 &  13.98 &   .21 &   .32 &   .47 &   .00 &  54 &   2.350\phantom{$^{\ast\ast}$} &  2.332 & 0.037 \nl 
U7387  & W  &  6 &  13.59 &   .36 &  1.44 &   .87 &   .07 &  90 &   2.411$^{\ast\ast}$ &  2.400 & 0.045 \nl 
N4289  & W  &  4 &  13.05 &   .36 &  1.48 &   .87 &   .02 &  90 &   2.555\phantom{$^{\ast\ast}$} &  2.554 & 0.032 \nl 
V 737  & W  &  7 &  14.44 &   .24 &   .59 &   .66 &   .03 &  70 &   2.215\phantom{$^{\ast\ast}$} &  2.207 & 0.047 \nl 
N4376  & Z  &  6 &  13.51 &   .21 &   .31 &   .46 &   .00 &  53 &   2.236\phantom{$^{\ast\ast}$} &  2.250 & 0.031 \nl 
U7522  & W  &  5 &  13.02 &   .35 &  1.42 &   .87 &   .03 &  90 &   2.495$^{\ast\ast}$ &  2.486 & 0.037 \nl 
N4405  & Z  &  5 &  12.70 &   .21 &   .29 &   .45 &   .03 &  51 &   2.674\phantom{$^{\ast\ast}$} &  2.236 & 0.038 \nl 
N4420  & W  &  5 &  12.38 &   .22 &   .48 &   .56 &   .00 &  63 &   2.344\phantom{$^{\ast\ast}$} &  2.338 & 0.020 \nl 
N4423  & Z  &  7 &  12.93 &   .29 &  1.08 &   .87 &   .00 &  87 &   2.233\phantom{$^{\ast\ast}$} &  2.232 & 0.065 \nl 
U7579  & W  &  6 &  14.24 &   .30 &  1.09 &   .87 &   .00 &  87 &   2.377\phantom{$^{\ast\ast}$} &  2.355 & 0.050 \nl 
N4455  & Y  &  5 &  12.23 &   .27 &   .81 &   .76 &   .06 &  77 &   2.188\phantom{$^{\ast\ast}$} &  2.098 & 0.079 \nl 
N4480  & W  &  3 &  12.73 &   .22 &   .41 &   .83 &   .01 &  63 &   2.619\phantom{$^{\ast\ast}$} &  2.563 & 0.023 \nl 
I3474  & W  &  7 &  13.61 &   .33 &  1.30 &   .87 &   .00 &  90 &   2.179\phantom{$^{\ast\ast}$} &  2.150 & 0.077 \nl 
U7697  & Y  &  5 &  13.87 &   .31 &  1.09 &   .87 &   .12 &  90 &   2.326\phantom{$^{\ast\ast}$} &  2.304 & 0.056 \nl 
N4527  & W  &  3 &  10.79 &   .23 &   .57 &   .98 &   .02 &  72 &   2.606\phantom{$^{\ast\ast}$} &  2.592 & 0.015 \nl 
N4533  & W  &  6 &  13.48 &   .30 &  1.13 &   .87 &   .00 &  90 &   2.254$^{\ast\ast}$ &  2.239 & 0.064 \nl 
N4536  & W  &  5 &  10.63 &   .23 &   .52 &   .59 &   .00 &  65 &   2.566$^{\ast\ast}$ &  2.548 & 0.016 \nl 
N4544  & S  &  5 &  13.20 &   .26 &   .75 &   .72 &   .05 &  75 &   2.305\phantom{$^{\ast\ast}$} &  2.314 & 0.044 \nl 
N4713  & Y  &  5 &  11.89 &   .21 &   .30 &   .46 &   .00 &  52 &   2.356\phantom{$^{\ast\ast}$} &  2.355 & 0.049 \nl 
U8114  & Y  &  5 &  14.62 &   .23 &   .55 &   .61 &   .00 &  67 &   2.182\phantom{$^{\ast\ast}$} &  2.176 & 0.047 \nl 
N4758  & Y  &  3 &  12.67 &   .27 &   .87 &  1.27 &   .01 &  90 &   2.297\phantom{$^{\ast\ast}$} &  2.282 & 0.059 \nl 
U8032  & Y  &  5 &  13.62 &   .27 &   .89 &   .80 &   .00 &  80 &   0.000$^{\dagger}$\phantom{$^{\ast}$} &  0.000 & 0.000 \nl 
I3881  & Y  &  5 &  12.66 &   .29 &   .93 &   .83 &   .12 &  82 &   2.318\phantom{$^{\ast\ast}$} &  2.310 & 0.052 \nl 
N4808  & Y  &  5 &  11.78 &   .23 &   .57 &   .61 &   .00 &  68 &   2.467\phantom{$^{\ast\ast}$} &  2.391 & 0.022 \nl 
U8085  & Y  &  3 &  13.71 &   .27 &   .87 &  1.27 &   .04 &  90 &   2.387\phantom{$^{\ast\ast}$} &  2.364 & 0.049 \nl 
\enddata

\tablecomments{$^1$ interacting pair N4567/8 (VV219)\hfill \break
$^2$ N4411A: Fouqu\'e et al.~(1990) give $i=30^{\circ}$\hfill \break
$^3$ peculiar galaxy (Binggeli et al.~1985)\hfill \break
$^4$ N4571: Fouqu\'e et al.~(1990) give $i=30^{\circ}$\hfill \break}
\end{deluxetable}

\clearpage
\begin{deluxetable}{ccccccccc}
\tablecaption{Comparison of TF parameters and the TF distance to the Virgo
cluster for different sets of input\hfill \break  parameters [cf.~eq.~(1)]}

\tablehead{
solution & slope & 
\multicolumn{2}{c}{intercept at $\log w=2.3$} &
\multicolumn{2}{c}{dispersion} & $n_{\rm Virgo}$ & 
$(m-M)_{\rm Virgo}$ & error \\
 & & \multicolumn{2}{c}{\hrulefill} &
\multicolumn{2}{c}{\hrulefill} & &
 & \\
 &  & Virgo & calibrators &
 Virgo & calibrators & & mag & mag \\
 (1) & (2) & (3) & (4) & (5) &
 (6) & (7) & (8) & (9)
}

\startdata
 1 & $-6.969$ &  13.13 & $-18.37$ &  0.57 & 0.44 & 49 & {\bf 31.50} & 0.09 \nl
 2 & $-7.071$ &  13.33 & $-18.23$ &  0.61 & 0.48 & 49 & {\bf 31.57} & 0.08 \nl
 3 & $-7.152$ &  13.13 & $-18.35$ &  0.60 & 0.43 & 67 & {\bf 31.48} & 0.07 \nl
 4 & $-6.551$ &  13.06 & $-18.94$ &  0.65 & 0.44 & 35 & {\bf 32.00} & 0.11 \nl
 5 & $-7.311$ &  13.11 & $-18.51$ &  0.62 & 0.39 & 49 & {\bf 31.63} & 0.10 \nl
 6 & $-7.371$ &  13.34 & $-18.40$ &  0.65 & 0.44 & 49 & {\bf 31.74} & 0.09 \nl
 7 & $-7.454$ &  13.09 & $-18.51$ &  0.65 & 0.38 & 67 & {\bf 31.60} & 0.08 \nl
 8 & $-6.925$ &  13.15 & $-18.39$ &  0.56 & 0.43 & 49 & {\bf 31.54} & 0.09 \nl
\hline
\multicolumn{2}{l}{mean over 1-7\hfill} & & & & & & {\bf 31.65$\pm$ 0.17} &  \nl
\enddata
\tablecomments{Parameters for the different solutions:
\begin{enumerate}
\item Magnitudes and $A^i$, $A^0$ from RC3, $\log w$ from LEDA
\item Magnitudes and from $A^i$, $A^0$ from RC3, $\log w$ from Huchtmeier and Richter (1989)
\item Like 1, but including all members of subclusters A+B with $i>30^{\circ}$
\item Like 1, but using the X-ray sample (excluding seven galaxies in the SW corner)
\item Magnitudes and $A^0$ from RC3, $A^i$ from RSA, $\log w$ from LEDA
\item Magnitudes and $A^0$ from RC3, $A^i$ from RSA, $\log w$ from Huchtmeier and Richter (1989)
\item Like 5, but including all members of subclusters A+B with $i>30^{\circ}$
\item Magnitudes and $A^i$ from RC3, $A^0=0$, $\log w$ from LEDA
\end{enumerate}
}
\end{deluxetable}

\clearpage
\begin{deluxetable}{lrrrrc}
\tablewidth{0pc}
\tablecaption{A comparison of Cepheid and Tully-Fisher distances (Table 4, solution 1)}
\tablehead{
\colhead{galaxy} & \colhead{$v_{\rm LG}$} & \colhead{$(m-M)_{\rm Cepheid}$} & \colhead{$(m-M)_{\rm TF}$} &
\colhead{$\Delta (m-M)$} & \colhead{subcluster}}

\startdata
N4321 & 1464 & $31.04\pm 0.26$ & $31.21\pm 0.40$ & $-0.17 \pm 0.50$ & A \nl
N4496A${^1)}$ & 1568 & $31.13\pm 0.10$ & $30.67\pm 0.40$ & $\phantom{-}0.46\pm 0.44$ & W \nl
N4536 & 1646 & $31.11\pm 0.05$ & $30.72\pm 0.40$ & $\phantom{-}0.39\pm 0.43$  & W \nl
N4639 & 860 & $32.00\pm 0.23$ & $32.53\pm 0.40$ & $-0.53\pm 0.49$ & A \nl
\enddata
\tablecomments{\small ${^1)}$ The magnitude may be disturbed by N\,4496B}
\end{deluxetable}

\clearpage
\begin{deluxetable}{lr}
\tablewidth{0pc}
\tablecaption{External errors of the Virgo cluster modulus}
\tablehead{
\colhead{error source} & \colhead{error}}
\startdata
zeropoint of the Cepheid P-L relation & 0\fm 10 \nl
adopted slope of the TF relation & 0\fm 10 \nl
precepts of weighting the data & 0\fm 08 \nl
weight of the X-ray sample & 0\fm 10 \nl
correction for internal absorption & 0\fm 10 \nl
intrinsic differences between calibrators and Virgo galaxies & 0\fm 05 \nl 
inclination difference between calibrators and Virgo galaxies & 0\fm 05 \nl
\hline
estimated total external error & 0\fm 23 \nl
\enddata
\end{deluxetable}

\clearpage
\begin{deluxetable}{lccccc}
\tablewidth{0pc}
\tablecaption{The values of $H_0$ calculated for subclusters A and B
separately}
\tablehead{
\colhead{} & \colhead{$n$} & \colhead{$v_{220}$} & \colhead{$(m-M)$}
& \colhead{$r$ (Mpc)} & \colhead{$H_0$}}
\startdata
A & 34 & $1164\pm67$ & $31.35\pm0.1$ & $18.6\pm 0.9$ & $63\pm4$ \nl
B & 15 & $1111\pm76$ & $31.81\pm0.1$ & $23.0\pm 1.6$ & $48\pm5$ \nl
\enddata
\end{deluxetable}